\documentclass[amsmath,12pt,amssymb,preprint,prd,aps,nofootinbib]{revtex4}
\usepackage{amsfonts} %\usepackage{citesort}
\usepackage{graphicx} % Include figure files
\usepackage{epsfig}
\usepackage{multirow}
\usepackage{bm}% bold math
\usepackage{subfig}
\begin{document}
\title{Spectral properties of $\omega$, $\rho$ and $A_1$ mesons in hot magnetized matter: effects of (inverse) magnetic catalysis}

\author{Pallabi Parui}
\email{pallabiparui123@gmail.com}
\author{Amruta Mishra}
\email{amruta@physics.iitd.ac.in}  
\affiliation{Department of Physics, 
Indian Institute of Technology, Delhi, Hauz Khas, New Delhi - 110016}
\begin{abstract}
In-medium masses of the light vector $\omega$, $\rho$ and axial-vector $A_1$ mesons are studied in the magnetized nuclear matter, accounting for the effects of (inverse) magnetic catalysis at finite temperature. The in-medium partial decay widths for the $A_1\rightarrow \rho \pi$ channels are studied from the in-medium masses of the initial and the final state particles, by applying a phenomenological Lagrangian to account for the $A_1\rho\pi$ interaction vertices. The masses are calculated within the QCD sum rule framework, with the medium effects coming through the light quark ($\sim \langle \bar{q}q \rangle$) and the scalar gluon condensates ($\sim \langle G^2 \rangle$), as well as the light four-quark condensate ($\sim \langle \bar{q}q\rangle^2 $). The condensates are calculated within the chiral $SU(3)$ model in terms of the medium modified scalar fields: isoscalar $\sigma$, $\zeta$, isovector $\delta$ and the dilaton field $\chi$. The effects of magnetic fields are incorporated through the magnetized Dirac sea contribution as well as the Landau energy levels of protons and anomalous magnetic moments (AMMs) of the nucleons at finite temperature nuclear matter. The effects of temperature are included through the Fermi distribution functions in the number ($\rho_{p,n}$) and scalar ($\rho^s_{p,n}$) densities of nucleons within the chiral effective model framework. The incorporation of the magnetic field through the Dirac sea of nucleons lead to an enhancement (reduction) of the light quark condensates with magnetic field, give rise to the phenomenon of magnetic (inverse) catalysis. The effects of (inverse) magnetic catalysis at finite temperature nuclear matter are studied on the spectral functions and production cross-sections of the neutral $\rho$ and $A_1$ mesons. This may affect the production of the light vector and axial-vector mesons in the peripheral heavy-ion collision experiments, where estimated magnetic field is very large at the early stages of collisions with very high temperature.     
\end{abstract}
\maketitle
\section{Introduction}
\label{sec1}
The study of the in-medium spectral properties (masses, decay widths etc.) of hadrons under extreme conditions of density and/or temperature, has become a very important topic of research in the strong interaction physics. The study is of great relevance in the context of high energy heavy-ion collision experiments. In the peripheral ultra relativistic heavy-ion collision experiments, very large magnetic fields have been estimated, for e.g., at RHIC in BNL, LHC in CERN \cite{kharzeev}-\cite{tuchin}. Thus, the in-medium study of hadrons in the presence of an external strong magnetic field attracts a lot of research interest in this area. The large number difference between the neutrons and protons of the heavy colliding nuclei leads to the incorporation of the effects of isospin asymmetry in the study of hadronic properties.

%The light axial-vector meson $A_1$ has isospin $I=1$ and parity, charge conjugation quantum numbers of $J^{PC}=1^{++}$, which for the light vector meson $\rho$ are $I^G(J^{PC})=1^+(1^{--})$. 
In-medium properties of the light vector mesons may affect the low mass dilepton production in the heavy-ion collision experiments \cite{rap}. The heavy leptonic decay of $\tau \rightarrow \nu_{\tau}+X$, indicates the coupling of the hadronic state $X$ to an axial-vector current. Its decay fraction to the three pion states through the intermediate state of $\rho\pi$ with respect to the total width shows that, the dominant decay mode of the $A_1$ meson is the partial ($s$-wave) $\rho\pi$ mode \cite{tao}. The study of the in-medium spectral properties of the $A_1$ meson is important in the context of partial restoration of chiral symmetry and to provide valuable information to further experimental studies of the axial-vector meson state. Its coupling to the $\pi$ meson $f_{\pi}$, is defined in the usual way $\langle0|\bar{u}\gamma_{\mu}\gamma_5d|\pi\rangle=if_{\pi}P_{\mu}$ \cite{B196}. The fundamental symmetry of Quantum chromodynamics (QCD) is the chiral symmetry, which is spontaneously broken at low densities and low temperatures due to the non-zero expectation values of the chiral condensates. The formation of the quark and gluon condensates in QCD vacuum leads to the generation of hadron masses. The spontaneous chiral symmetry breaking effect induces mass splittings between the opposite parity states in the hadron spectra, e.g., $\pi$-$\sigma,$ $\rho$-$A_1$. \\
The condensates are expected to change with temperature and/or density, and also with the magnetic field. The values tend to decrease with baryon density in the strange hadronic matter at different values of strangeness $f_s=0,\ 0.3,\ 0.5$ and isospin asymmetry parameter $\eta=0,\ 0.5$ \cite{91}. The phenomena of enhancement (reduction) of the light quark condensates with magnetic field is called magnetic (inverse) catalysis \cite{Shovkovy, elia, kharzeevmc}. In the literature, there are few studies related to the effects of (inverse) magnetic catalysis on the hadron properties in magnetized nuclear matter. There are several studies in the context of quark matter with a background magnetic field using the framework of Nambu-Jona-Lasinio model \cite{Preis, menezes, ammc, lemmer, guinjl}. The effects of the magnetized Dirac sea have been studied on the nuclear matter phase transition using the Walecka model and an extended linear sigma model \cite{haber}. %An increasing mass of nucleon as a function of the magnetic field is obtained, at zero baryon density and zero anomalous magnetic moments of the nucleons. %The effect of magnetic catalysis is observed indirectly through the scalar field dependency of the nucleon mass $M_N=m_N-g_{\sigma}\sigma$, where the scalar fields are proportional to the light quark condensates (with $m_N$ the vacuum mass of nucleon and $g_{\sigma}$ is the $\sigma$ meson-nucleon coupling constant). 
In ref. \cite{arghya}, the effects of (inverse) magnetic catalysis have been studied using the weak-field expansion of the fermion propagator incorporating its anomalous magnetic moment, to evaluate the nucleonic one-loop self energy functions in the hot magnetized nuclear matter. %An increase of the nucleon mass with magnetic field is observed at zero density within the Walecka model, in addition to some significant contribution from the anomalous magnetic moments of the Dirac sea of nucleons.
%At finite temperature, accounting for the anomalous magnetic moments of nucleons, an inverse magnetic catalysis effect \cite{balicm} was observed on the critical temperature of the vacuum to nuclear matter phase transition. The opposite behavior is obtained when the nucleons AMMs are taken to be zero \cite{arghya}. 
At very high temperatures and high densities, the chiral symmetric phase is expected to be (partially) restored, which can be inferred from its observable consequence on the hadron spectrum. The dilepton data, as measured in the relativistic heavy-ion collisions, provided evidence on the in-medium spectral changes of $\rho$ meson, while there is very little access to the experimental evidence on the spectral changes of $A_1$ meson. Therefore, to investigate on the chiral symmetry restoration from the in-medium spectral changes of the $\rho$ meson \cite{hohler}, a theoretical investigation of the same is required on the $A_1$ meson spectrum. Towards this aim, sum rules serve as good non-perturbative tool to connect the hadronic spectral properties directly to the QCD vacuum condensates. The medium modifications of the spectral properties of hadrons may serve as good evidence to the in-medium nature of QCD vacuum condensates.

There have been QCD sum rule studies on the spectral density functions induced by the axial-vector current \cite{B196, shifman448}. %In these works, Borel sum rules have been constructed for the axial-vector current and divergence of the current, which lead to the same mass value for $A_1^0$ at a particular value of the continuum threshold, $s_0$ \cite{ b283}. 
The spectral functions of $\rho$ and $A_1$ mesons have been analyzed in vacuum using the hadronic models and have been tested further with QCD sum rules \cite{rapp}. The Weinberg-type sum rules \cite{weinberg} for $\rho$ and $A_1$ mesons (in the exact chiral limit), have been studied at zero and finite temperature \cite{49}. %A thorough discussion were given on the possible relations between the chiral symmetry restoration and the various types of finite temperature sum rules. 
In ref. \cite{hatsudab394}, finite temperature ($ T\neq $ 0) study of $\rho$, $\omega$ and $A_1$ mesons have been performed by using Borel sum rule. %The study has shown a drop in the $A_1$ mass with temperature. 
The effects of temperature have been incorporated through the thermal average of the local operators in the operator product expansion, which leads to the non-vanishing values of Lorentz non-scalar operators that would have otherwise been zero at $T=0$. The contributions of the scalar four-quark condensates are observed to be significant on these meson properties. In ref. \cite{leupold}, the Breit-Wigner parametrization for the $\rho, A_1$ spectral functions have been used and the constraints of QCD sum rules have been investigated on their masses and decay widths, in vacuum and at finite nuclear matter density. The coupling of $\pi$ meson to the axial-vector current has been considered by adding a $\delta$-function peak at $m_{\pi}$ to the correlator of axial-vector channel. In ref. \cite{kwon}, the parity-mixing ansatz including the finite widths of $\rho, A_1$ spectra has been considered at finite temperature in the context of finite energy sum rules. The mass of the $A_1$ meson is reported to drop with rising temperature, indicating the expected tendency of $\rho$ -$A_1$ mass degeneracy nearby the critical temperature $T_c$ of chiral symmetry restoration. The temperature dependencies of the meson-nucleon coupling constants for $\rho$ \cite{sahin} and $A_1$ \cite{sahin2}, have been investigated using the soft-wall AdS/QCD model with thermal dilaton field. An interaction Lagrangian of the $\rho(A_1)$-nucleon interaction along with the thermal dilaton field system has been constructed in the bulk of space-time, to form the integral representation of the $g_{A_1NN}(g_{\rho NN})$ coupling. The mesons and nucleons profile functions are applied to the model. The AdS/CFT correspondence states the equivalence of two different physical theories of gravity (in the bulk of AdS space-time) and a quantum field theory (on the boundary of this space-time). There is a strong-weak duality which is applied to describe the low-energy phenomena of QCD (theory of strong interactions), and is called AdS/QCD. There are two approaches in AdS/QCD used to study the hadron properties in nuclear medium, which is expected to create in the relativistic heavy ion collision experiments. The modified thermal soft-wall model has been used to incorporate the effects of temperature through the thermal dilaton field, which is also related to the chiral condensate. 

In ref. \cite{100}, the effects of magnetic field have been studied on the in-medium masses of the light vector mesons ($\rho, \omega, \phi$) in magnetized nuclear matter using the QCD sum rule method. In-medium masses are obtained by incorporating the medium effects through the scalar fields $\sigma, \zeta, \delta, \chi$, calculated within the chiral $SU(3)$ model. In the absence of an external magnetic field, masses of these mesons have been studied in the strange hadronic matter using the sum rule approach \cite{91}, in terms of the light quark (up to the scalar four quark condensates) and the scalar gluon condensates calculated within the chiral model framework. In ref. \cite{100}, contribution of an external magnetic field has been incorporated through the Landau quantization of protons and the anomalous magnetic moments of the nucleons. In our present study, contributions of the magnetized Dirac sea are incorporated to study the effects of (inverse) magnetic catalysis on the masses of $\rho$, $\omega$ and $A_1$ mesons as well as the spectral function and production cross-sections of $\rho^0, A_1^0$ mesons in hot and dense nuclear matter, in an external magnetic field.\\
The present paper is organized as follows: in sec.\ref{sec2}, the chiral effective model is briefly discussed to find the in-medium light quark and the scalar gluon condensates. In sec.\ref{sec3}, QCD sum rule approach is presented to calculate the masses of the light vector and axial-vector mesons under study. Sec.\ref{sec4}, describes the phenomenological Lagrangian formulation to find the hadronic decays of the $A_1$ and $\rho$ mesons. Sec.\ref{sec5} contains the formalism of the Breit-Wigner spectral functions and production cross-sections of the neutral $\rho$ and $A_1$ mesons incorporating the medium effects through their respective mass and decay widths.  The results of the present investigation are discussed in sec.\ref{sec6}. Finally, sec.\ref{sec7}, summarizes the findings of the present work.  

\section{The chiral effective Model}
\label{sec2}
The in-medium masses of the light vector and axial-vector mesons within the QCD sum rule approach are obtained through the light quark and scalar gluon condensates, which are calculated using a chiral effective model framework. The effective chiral model is based on the non-linear realization of chiral $SU(3)_L\times SU(3)_R$ symmetry \cite{coleman, weinberg1, bardeen} and the broken scale-invariance of QCD \cite{papa, 69, zschi}. A scale-invariance breaking logarithmic potential in the scalar dilaton field $\chi$ \cite{sech, ellis} within the effective Lagrangian simulates QCD gluon condensate. %The Lagrangian density of the chiral $SU(3)$ model can be generalized as \cite{papa}, \\
%\begin{equation}
   %\mathcal{L}=\mathcal{L}_{kin}+\mathcal{L}_{BM}+\mathcal{L}_{vec}+\mathcal{L}_0+\mathcal{L}_{scale-break}+\mathcal{L}_{SB}+\mathcal{L}_{mag} 
%\end{equation}
%$\mathcal{L}_{kin}$ is the kinetic energy of the baryons and the mesons degrees of freedom; $\mathcal{L}_{BM}$ represents the baryon-mesons (both spin-0 and spin-1 mesons) interactions; $ \mathcal{L}_{vec}$ contains the quartic self-interactions of the vector mesons and their couplings with the scalar ones; $\mathcal{L}_0$ incorporates the spontaneous chiral symmetry breaking effects via meson-meson interactions; $\mathcal{L}_{scale-break}$ is the scale symmetry breaking logarithmic potential; and $\mathcal{L}_{SB}$ is the explicit symmetry breaking term; finally
The general Lagrangian density of the model contains the kinetic energies of the baryons and mesons, along with the baryon-meson (both spin-0 and spin-1) interactions. Some of which generate the mass of the baryons (due to their couplings with the scalar mesons fields). The meson-meson interactions give rise to their mass and dynamics in the effective model Lagrangian. The spontaneous symmetry breaking and explicit symmetry breaking of QCD are incorporated within the chiral $SU(3)$ model Lagrangian accordingly \cite{papa}. The scale-symmetry breaking logarithmic potential simulates the scalar gluon condensates in the model. %The effects of an external magnetic field on the nuclear matter are incorporated due to the electric charge of the protons and finite anomalous magnetic moments of the nucleons \cite{p97, am98, am981, prakash, brod, wei, mao}. %charged and neutral baryons in the nuclear medium are incorporated through  
%\begin{equation}
%\mathcal{L}_{mag}=-\frac{1}{4}F_{\mu\nu}F^{\mu\nu}-e_i{\bar{\psi}}_i\gamma_\mu A^\mu\psi_i-\frac{1}{4}\kappa_i\mu_N{\bar{\psi}}_i\sigma^{\mu\nu}F_{\mu\nu}\psi_i
%\end{equation}
%where,  $\psi_i$ is the baryon field operator for $i^{th} ( i = p, n)$ baryon with electric charge $e_i$, in the nuclear matter. The parameters, $\kappa_p (i=p)= 3.5856$ and $\kappa_n (i=n) = -3.8263$, are the gyromagnetic ratio corresponding to the anomalous magnetic moments (AMMs) of the proton and the neutron, respectively \cite{wei, mao}.
In the magnetized nuclear medium, the magnetic field contributions are coming through the Landau energy levels of protons and anomalous magnetic moments of the nucleons in the Fermi sea \cite{p97, am98, am981, prakash, brod, wei, mao}.

In some earlier works on the in-medium vector meson properties using Quantum Hadrodynamics model, the dominant contribution to the mass shifts are originated from the vacuum polarization effects in the baryon (nucleon) sector.
This significant drop can not be obtained while considering the mean-field approximation only. The in-medium properties of vector mesons and their observable effects on the low mass dilepton spectra have been studied extensively in the literature accounting for the quantum correction (through relativistic Hartree approximations) effects within the Walecka model. It is therefore interesting in the context of modern day experimental facilities at RHIC, LHC, where at the peripheral heavy ion collisions ultra strong transient electromagnetic field is expected to be produced at the very early stages of the collisions, which eventually decay with time in the produced medium. Depending on the collision geometry, the effect of an uniform magnetic field background in a less dense medium can be considered and the quantum corrections in the nucleonic sector (as in the low density matter the possible constituents can be protons and neutrons) should be taken into account.
 
The propagation of a particle through an external medium at finite density and/or temperature lead to the modifications to its field theoretical propagator. The same is also true when the particle is going through an external magnetic background. In the formalism of the relativistic Hartree approximation (RHA), the effective interaction of a fermion e.g., nucleon in a medium is taking into account through the one-loop self-energy function derived using the Feynman tadpole diagrams. The diagrams are corresponding to the scalar and vector mesons interactions with the nucleons. The tadpole diagrams, in turn, correspond to the second-order contributions to the various Green’s functions, e.g., fermion propagators. In the non-central ultra-relativistic heavy ion collisions, there have been estimation of very strong electromagnetic fields at the very early stages of the collisions. The strength of the produced fields may depend on the various collision parameters and collision geometry. However, the magnetic field at the freeze-out hyper-surface (FOHS) is estimated to be of the order of $0.07$ Ge$V^2$ which is around $4 m_{\pi}^2$ \cite{fohs}. In order to obtain a more complete in-medium effects, the effect of finite temperature is also accounted for in the present study. As it will be discussed further in the results and discussion section, the solutions of the scalar fields within the present model incorporating the magnetized Dirac sea contributions, are obtained up to a magnetic field strength of 3.9$m_{\pi}^2$ in the vacuum and almost around 9$m_{\pi}^2$ at the nuclear matter saturation density $\rho_0$, which is nearby the field strength at the FOHS. In this process, the one-loop baryons i.e., protons and neutrons are completely dressed in the sense that their effective interacting propagator is considered at the given medium while computing the loop integrals. The effective interacting propagator of fermion is obtained from the perturbative expansion in the Dirac equation in presence of an external magnetic field with finite anomalous magnetic moments (AMMs) of the nucleons. The weak-field expansion is considered in the present study by keeping up to the second order contributions in magnetic field and AMMs of the Dirac sea of nucleons. The method of dimensional regularization is applied to obtain the simplified form of the divergent integrals involved in the one-loop self energy functions. The ultraviolet divergence resulting from the pure vacuum term is neglected using the mean-field approximation. However, the other divergent contribution which is extracted from the pole of the gamma function is treated using the $\overline{MS}$ renormalization scheme. In ref.\cite{arghya}, the above procedure has been implemented to study the vacuum to nuclear matter phase transition at finite temperature nuclear matter within the Walecka model by considering $\sigma-N$ interaction in the one-loop scalar self energy function of nucleons. In our present work, the formalism is extended to apply in the context of chiral $SU(3)$ model, which incorporates the $\sigma-N$, $\zeta-N$ and $\delta-N$ interactions in the scalar one-loop self energy functions. The contribution of the anomalous magnetic moments of the Dirac sea of nucleons is considered in the weak field expansion of the magnetized fermion propagator. Thus, the contribution of the magnetized Dirac sea is incorporated through the summation of the nucleonic tadpole diagrams corresponding to their interactions with the scalar meson fields $\sigma$, $\zeta$ and $\delta$ in the chiral $SU(3)$ model Lagrangian.
\begin{align}
    \mathcal{L}_{BX} & =\sum_{i=p,n} \bar{\psi}_{i}(g_{\sigma i}\sigma+g_{\zeta i}\zeta+g_{\delta i}\delta)\psi_{i}.
\end{align}
Where, $\sigma$, $\zeta$  are the non-strange and strange scalar-isoscalar fields, respectively and $\delta$ is the scalar-isovector field. The coupling parameters $g_{\sigma i},\ g_{\zeta i}, \ g_{\delta i}$  corresponding to the $i^{th};\ i=p,n$, baryon are fitted from the empirically known masses of the baryons. The one-loop self energy functions of the Dirac sea of nucleons are thus evaluated using the weak-field expansion of nucleonic propagators (up to second order in magnetic field). The effects of AMMs of the Dirac sea of nucleons are also accounted for in this study. The weak-field approximation of fermion propagator leads to a reliable solutions of the self-consistent scalar fields equations in the magnetized nuclear matter, till the magnetic field strength of $0.04$ Ge$V^2$ or around 2-3 $m_{\pi}^2$ within the Walecka model \cite{arghya}. In the present work's chiral $SU(3)$ model, the similar field expansion of the propagators of Dirac nucleons is considered to find the magnetized vacuum corrections in the nuclear matter context. The self-consistent scalar fields equations for $\sigma$, $\zeta$, $\delta$ and $\chi$ including both Fermi and Dirac sea effects in the magnetic nuclear matter give rise to a feasible solution up to a magnetic field strength which is consistent with that of the freeze-out hypersurface for hadronic cooling zone. As there is a finite chance of the in-medium hadron properties to be affected by a low-strength magnetic field in this regime. In this sense the weak-field expansion is considerable for a qualitative estimate of the hadron properties in a magnetized matter. Beside this the degree of divergence involved in the loop integral would also be out of control if higher order terms in the propagator expansion were taken into account. The nucleonic self energy functions contribute to the scalar densities of nucleons \cite{chrmmc, dmc, bmc}, which then give impact to the scalar fields coupled equations of motion. The scalar meson fields are treated as classical, whereas the nucleons as quantum fields in the evaluation of the magnetized Dirac sea contribution. 
% \\ The concept of broken scale-invariance of QCD is incorporated in the chiral model at the tree level through a logarithmic potential in the scalar dilaton field $\chi$ \cite{sech, ellis} as,  
%\begin{equation}
 %   \mathcal{L}_{scale-break}=-\frac{1}{4}\chi^4\ln\left(\frac{\chi^4}{\chi_0^4}\right) + \frac{d}{3}\chi^4\ln\left(\left(\frac{(\sigma^2-\delta^2)\zeta}{\sigma_0^2\zeta_0}\right)\left(\frac{\chi}{\chi_0}\right)^3\right).
%\end{equation}
% The Lagrangian parameter $d$ is chosen to be 0.064 \cite{papa, arindam79}. The subscript-$0$ in the fields indicate their respective vacuum expectation values.
The scale-invariance breaking phenomena of QCD leads to the trace anomaly of QCD, i.e., non-zero value for the trace of the energy-momentum tensor in QCD, which in the limit of finite light quark masses $m_i (i=u,d,s)$ become \cite{100, cohen}  
\begin{equation}
    \langle T^{\mu}_{\mu}  \rangle= \sum_{i=u,d,s}\langle m_i\bar{q}_i q_i\rangle + \langle \frac{\beta_{QCD}}{2g}G_{\mu\nu}^a G^{a\mu\nu}\rangle.
    \label{emttr1}
\end{equation}
In eq.(\ref{emttr1}), $G_{\mu\nu}^a$ is the gluon field strength tensor of QCD. The trace of the energy momentum tensor within the model can be obtained from the Lagrangian density terms containing the field $\chi$ \cite{91,arvind82, heide} as
\begin{equation}
    \langle\theta_{\mu}^{\mu}\rangle = \chi\frac{\partial\mathcal{L}}{\partial\chi} -4\mathcal{L}=-(1-d)\chi^4
    \label{emttr2}
\end{equation}
The first term in eq.(\ref{emttr1}) corresponds to the explicit chiral symmetry breaking term in QCD 
\begin{equation}
    \mathcal{L}_{SB}^{QCD}=- Tr[diag(m_u\bar{u}u,m_d\bar{d}d,m_s\bar{s}s)],
    \label{sb1}
\end{equation}
which in the chiral $SU(3)$ model under the mean-field approximation is written as \cite{91} 
\begin{equation}
\mathcal{L}_{SB}= Tr\left[ diag\left( -\frac{1}{2}m_\pi^2 f_\pi(\sigma+\delta),  -\frac{1}{2}m_\pi^2 f_\pi(\sigma-\delta), \left(\sqrt{2}m_k^2f_k-\frac{1}{\sqrt{2}}m_\pi^2 f_\pi\right)\zeta\right)\right]
\label{sb2}
\end{equation}
Comparing eqs. (\ref{sb1}) and (\ref{sb2}), the light quark condensates are related to the scalar fields $\sigma$, $\zeta$ and $\delta$ as 
\begin{align}
    m_u\langle\bar{u}u\rangle & = \frac{1}{2}m_\pi^2 f_\pi(\sigma+\delta)
    \label{uu},\\
    m_d\langle\bar{d}d\rangle & =\frac{1}{2}m_\pi^2 f_\pi(\sigma-\delta)
    \label{dd}, \\
    m_s\langle\bar{s}s\rangle & = \left(\sqrt{2}m_k^2f_k-\frac{1}{\sqrt{2}}m_\pi^2 f_\pi\right)\zeta
    \label{ss}
\end{align}
From eqs. (\ref{emttr1}) and (\ref{emttr2}), one obtains the scalar gluon condensate as 
\begin{equation}
\sum_{i=u,d,s}m_i\langle\bar{q}_i q_i\rangle -\frac{9}{8} \left\langle \frac{\alpha_{s}}{\pi}G_{\mu\nu}^a G^{a\mu\nu}\right\rangle =-(1-d)\chi^4
     \end{equation}
In eq.(\ref{emttr1}), $\beta_{QCD}(g) = -\frac{N_cg^3}{48\pi^2} (11-\frac{2}{N_c} N_f )=-\frac{9\alpha_sg}{4\pi}$ is the QCD $\beta$ function at the one-loop level with $N_f=3$ flavors, $N_c=3$ colors of quarks and $\alpha_s=\frac{g^2}{4\pi}$; by using the expressions for $m_i\langle\bar{q_i}q_i\rangle$ $(i=u,d,s)$ from eqs. (\ref{uu})-(\ref{ss}), the scalar gluon condensate is given by
\begin{equation}
    \left\langle \frac{\alpha_{s}}{\pi}G_{\mu\nu}^a G^{a\mu\nu}\right\rangle =\frac{8}{9}\left[(1-d)\chi^4 + \left( m_\pi^2 f_\pi \sigma +\left(\sqrt{2}m_k^2f_k-\frac{1}{\sqrt{2}}m_\pi^2 f_\pi\right)\zeta\right) \right]
    \label{gg}
\end{equation}
The coupled equations of motion of the scalar fields $\sigma$, $\zeta$, $\delta$ and $\chi$ obtained from the chiral $SU(3)$ model Lagrangian are solved in the magnetized (asymmetric) nuclear matter at finite temperature, accounting for the effects of the Dirac sea under the mean-field approximation. In the coupled equations of motion \cite{arvind82}, the number and scalar densities of nucleons ($\rho_i, \rho^{s}_i; i=p,n$) incorporate the effects of density, temperature and magnetic fields as given below for the charged protons and neutral neutrons in hot magnetized nuclear matter, %\cite{rabhi, am107},
\begin{align}
    \rho_p & = \frac{|eB|}{2\pi^2}\sum_{\nu,\  s=\pm 1}\int_0^{\infty}dk_{||}\left( \frac{1}{e^{(E_p^*-\mu_p^*)/T}+1} - \frac{1}{e^{(E_p^*+\mu_p^*)/T}+1} \right) 
    \label{rhp},
    \end{align}
    \begin{equation}
     \rho_p^s  = \frac{|eB|m^*_p}{2\pi^2}\sum_{\nu,\ s=\pm1} \int_0^{\infty} \frac{dk_{||}}{\sqrt{k_{||}^2 + (\sqrt{m^{*2}_p+2\nu |eB|}+s\Delta_p)^2}} \left( \frac{1}{e^{(E_p^*-\mu_p^*)/T}+1} + \frac{1}{e^{(E_p^*+\mu_p^*)/T}+1} \right)   + \Delta \rho^s_p
     \label{rhsp},
     \end{equation}
     
     \begin{equation}
    \rho_n  = \frac{1}{2\pi^2}\sum_{s=\pm 1}\int_0^{\infty} dk_{\perp}  k_{\perp} \int_0^{\infty} dk_{||} \left( \frac{1}{e^{(E_n^*-\mu_n^*)/T}+1} - \frac{1}{e^{(E_n^*+\mu_n^*)/T}+1} \right)
    \label{rhn},
    \end{equation}
    \begin{multline}
    \rho_n^s  = \frac{1}{2\pi^2} \sum_{s} \int_0^{\infty} dk_{\perp} k_{\perp} \left( 1+ \frac{s\Delta_n}{\sqrt{m_n^{*2}+k_{\perp}^2}}\right)  \int_0^{\infty} dk_{||}\frac{m^*_n}{\sqrt{k_{||}^2 + (\sqrt{m^{*2}_n+ k_{\perp}^2} + s\Delta_n)^2}}\\ \times \left( \frac{1}{e^{(E_n^*-\mu_n^*)/T}+1} + \frac{1}{e^{(E_n^*+\mu_n^*)/T}+1} \right)  + \Delta \rho^s_n.
    \label{rhsn}
\end{multline}
In the above expressions, the term related to the anomalous magnetic moments of protons and neutrons is $\Delta_{i(i=p,n)}= -\frac{\kappa_i \mu_{N}B}{2}$, with $\kappa_p= 3.5856$ and $\kappa_n  = -3.8263$, are the gyromagnetic ratio corresponding to the anomalous magnetic moments (AMMs) of the proton and the neutron, respectively \cite{wei, mao}. The energy spectra of protons and neutrons in the magnetized nuclear matter are $E^*_p=\big[k_{||}^2 + (\sqrt{m^{*2}_p+2\nu |eB|}+s\Delta_p)^2\big]^{1/2}$, and $E^*_n = \big[k_{||}^2 + (\sqrt{m^{*2}_n+ k_{\perp}^2} + s\Delta_n)^2\big]^{1/2}$; $\mu^*_{i}$ denote the effective chemical potential of protons and neutrons ($i=p,n$) at finite temperature matter \cite{am81, am91}. In equations (\ref{rhsp}) and (\ref{rhsn}), the additional contribution $\Delta \rho^s_i;\ ({i=p,n})$ due to the magnetized Dirac sea effects can be related to the vacuum self-energy corrections in a background magnetic field as  
\begin{multline}
     - \frac{g_{Si}^2}{m_S^2}\Delta\rho^s_N \Big{|}_{S=\sigma, \zeta, \delta}= \Sigma^s_{vacuum} = \frac{g_{Si}^2}{4\pi^2m_S^2}\Bigg[ \frac{|eB|^2}{3m^*_i} + \Big(\Delta_p^2m^*_p + \Delta_n^2m^*_n -  |eB|\Delta_p \Big) \left\{\frac{1}{2}+ 2ln\left(\frac{m^*_i}{m_i}\right)\right\}
 \Bigg]
 \label{selfexpress}
\end{multline}
In equation (\ref{selfexpress}), the corresponding electric charge for proton should be used which is zero for neutron. In the magnetized nuclear matter, the number and scalar densities of charged fermions (protons) have the contributions of the Landau energy levels of protons. The number and scalar densities of both protons and neutrons incorporate the effects of the anomalous magnetic moments (AMMs) of the nucleons from the magnetized Fermi sea of nucleons \cite{p97, am98, wei, mao, rabhi}. The scalar densities of the nucleons incorporate an additional contribution ($\Delta \rho_i^s; i=p,n$) due to the magnetized Dirac sea, which is obtained by summing over the nucleonic tadpole diagrams corresponding to the scalar mesons ($\sigma$, $\zeta$ and $\delta$) and nucleons interactions within the chiral effective model. %The weak-field expansion of the fermionic propagator is considered in the self-energy calculations of nucleons with finite anomalous magnetic moments of the nucleons \cite{arghya}.
The Fermi distribution functions give rise to the effect of temperature on the solutions of the scalar fields ($\sigma$, $\zeta$, $\delta$ and $\chi$) through $\rho_{i}$ and $\rho^s_i;\ i =p,n$ \cite{rabhi}. The scalar fields equations of motion are thus solved self-consistently at the given values of the baryon density $\rho_B=\rho_p+\rho_n$, isospin asymmetry parameter $\eta=\frac{\rho_n -\rho_p}{2\rho_B}$, temperature $T$, and magnetic fields $|eB|$, taken into account the important effects of the magnetic field through Dirac sea.   %The increasing (decreasing) values of the light quark condensates with magnetic field represent the (inverse) magnetic catalysis, which are proportional to the scalar fields as given by eqs.(\ref{uu})-(\ref{ss}). Thus, the effects of (inverse) magnetic catalysis are studied on the in-medium masses of the $\rho,\ \omega$ and $A_1$ mesons, by using the QCD sum rule method as it is described in the next section.  

\section{QCD sum rule framework}
\label{sec3} 
 In the present section, the masses of the light vector mesons $\omega$, $\rho$ and the axial-vector meson $A_1$ are calculated using the QCD sum rule (QCDSR) approach. In-medium masses of these mesons are investigated in the magnetized nuclear matter at finite temperature, accounting for the effects of (inverse) magnetic catalysis through the magnetized Dirac sea. The in-medium masses are obtained in terms of the light quark condensates (up to the scalar four-quark condensate) and the scalar gluon condensate, which are calculated within the chiral $SU(3)$ model. The time-ordered current-current correlator is given by \cite{B196, shifman448}
\begin{equation}
     \Pi_{\mu\nu}(q)= i \int d^4x \  e^{iqx}\left< T[J_{\mu}(x), J_{\nu}(0)] \right>.  
\end{equation}
Where $'T'$ denotes the time-ordered product, and the symbol $'\langle\rangle'$ indicates the in-medium expectation value of the T-ordered product of currents. The current quark-bilinears of all the three axial-vector meson states (with $J^P=1^+$) are defined as 
\begin{equation}
    J^{(A_1^{+})}_{\mu} = \bar{d}\gamma_{\mu}\gamma_{5}u; \quad J^{(A_1^{-})}_{\mu} = \bar{u}\gamma_{\mu}\gamma_{5}d; \quad J^{(A_1^0)}_{\mu}=\frac{1}{2}(\bar{u}\gamma_{\mu}\gamma_{5}u-\bar{d}\gamma_{\mu}\gamma_{5}d)
\end{equation}
For the vector currents with $J^{P}=1^{-}$ they are given by 
\begin{equation}
    J^{(\rho^+)}_{\mu}= \bar{d}\gamma_{\mu}u; \quad J^{(\rho^-)}_{\mu} = \bar{u}\gamma_{\mu}d;\quad J^{(\omega,\ \rho^0)}_{\mu}= \frac{1}{2}(\bar{u}\gamma_{\mu}u\pm\bar{d}\gamma_{\mu}d)
\end{equation}
 The correlation function is written out in the following tensor structure \cite{shifman448, leupold, k1, urban}
 \begin{equation}
     \Pi_{\mu\nu}(q) = q_{\mu}q_{\nu} R(q^2) -g_{\mu\nu} K(q^2)
 \end{equation}
 The above expression is valid for mesons at rest. For the conserved vector currents, $K(q^2)=q^2R(q^2)$. As for the non-conserved axial-vector current this relation no longer holds and $R(q^2)$ has contributions from pseudoscalar mesons \cite{k1, shifman448, urban}. In principle, QCD sum rule can be carried out with either $R(q^2)$ or $K(q^2)$. Further studies in this work are based on $R(q^2)$ as the Borel transform with the other quantity is rather unstable and there will be no new information obtained by working with $K(q^2)$ along with $R(q^2)$ for the $A_1$ current \cite{leupold}. \\ 
 Among the two representations of QCDSR, the real part of the correlator $R(q^2)$ on the phenomenological side is related to its imaginary part via a dispersion relation \cite{91,k1} 
 \begin{equation}
       R_{phen.} (q^2) = \frac{1}{\pi}\int^{\infty}_0 ds\ \frac{ImR^{phen.} (s)}{(s-q^2)}.
       \label{pheno}
  \end{equation}
 Where $ImR^{phen.}(s)$ is called the spectral density, parametrized in terms of the hadronic resonance plus 
 perturbative continuum. On the other representation, the real part of $R(q^2)$ is expressed in the large space like region $(Q^2=-q^2)>>1$ $GeV^2$ by the Wilson's operator product expansion (OPE), given as \cite{hatsudab394,klingl}
\begin{equation}
    R_{OPE} (q^2=-Q^2) = \left( -c_0 \ln\left(\frac{Q^2}{\mu^2}\right) + \frac{c_1}{Q^2} + \frac{c_2}{Q^4} + \frac{c_3}{Q^6} + ...\right) 
    \label{ope}
\end{equation}
  In the above equation the operators up to dimension-6 are considered and the scale $\mu$ has been chosen as 1 GeV \cite{91,hatsudab394, klingl}. The first term in the OPE is a contribution from perturbative QCD. The coefficients $c_i$  $(i=1,2,3)$ of the subsequent terms contain QCD non-perturbative effects in terms of the light quark and scalar gluon condensates and some parameters from the QCD Lagrangian. The condensates are affected in presence of a medium. The effects of magnetic field through the Dirac sea lead to the significant changes in the condensates with magnetic field, which are of main concern of study on the in-medium spectral properties of light mesons. The coefficient $c_3$ is being associated with the light four-quark condensate and it is different corresponding to the different current quark-bilinears of the charged and neutral mesons with the same $J^{PC}$ quantum numbers. In Eq.(\ref{ope}), the $c_i$ $(i=0,1,2,3)$ coefficients for the light axial-vector meson $A_1$ are given by \cite{hatsudab394, leupold, k1, F1} 
\begin{equation*}
    c_0=\frac{1}{8\pi^2} \left( 1+\frac{\alpha_s}{\pi}\right ); \quad    c_1=-\frac{3}{8\pi^2}(m_u^2+m_d^2); \quad c_2=\frac{1}{24}\langle\frac{\alpha_s}{\pi}G^{\mu\nu}G_{\mu\nu}\rangle - \frac{1}{2}(m_u\langle  \bar{u}u\rangle+ m_d\langle \bar{d}d\rangle), 
\end{equation*}
%\begin{equation}
 %   c_1=-\frac{3}{8\pi^2}(m_u^2+m_d^2)
%\end{equation}
%\begin{equation*}
%    c_2=\frac{1}{24}\langle\frac{\alpha_s}{\pi}G^{\mu\nu}G_{\mu\nu}\rangle - \frac{1}{2}\langle m_u \bar{u}u+m_d\bar{d}d\rangle 
%\end{equation*}
and
\begin{equation*}
    c_3^{(A_1^0)} = \pi \alpha_s \times \frac{88}{81}\kappa_{1}\left(\langle\bar{u}u\rangle^2 + \langle\bar{d}d\rangle^2\right).
 \end{equation*}
    %-\frac{\pi}{2}\alpha_s\Bigg[\left\langle\left(\bar{u}\gamma_{\mu}\lambda^a u - \bar{d}\gamma_{\mu}\lambda^ad\right)^2\right\rangle   + \frac{2}{9} \bigg\langle\left(\bar{u}\gamma_{\mu}\lambda^a u + \bar{d}\gamma_{\mu}\lambda^ad\right) \times \\
%\left(\sum_{q=u,d,s}\bar{q}\gamma^{\mu}\lambda^a q\right)\bigg\rangle\Bigg]= \pi \alpha_s \times \frac{88}{81}\kappa_{1}\left(\langle\bar{u}u\rangle^2 + \langle\bar{d}d\rangle^2\right)
%\begin{equation*}
%    c_3^{(A_1^{\pm})}= \pi\alpha_s\times \kappa_{2} \left[ \frac{16}{81}\left(\langle\bar{u}u\rangle^2 + \langle\bar{d}d\rangle^2\right) + \frac{16}{9}\langle\bar{u}u\rangle\langle\bar{d}d\rangle \right]
%\end{equation*}
%\begin{multline}
%    c_3^{(A_1^{\pm})}=-\frac{\pi}{2}\alpha_s\Bigg[2\left\langle\left(\bar{u}\gamma_{\mu}\lambda^a d\right)\left(\bar{d}\gamma_{\mu}\lambda^a u\right)\right\rangle   + \frac{2}{9} \bigg\langle\left(\bar{u}\gamma_{\mu}\lambda^a u + \bar{d}\gamma_{\mu}\lambda^ad\right) \times \\
%\left(\sum_{q=u,d,s}\bar{q}\gamma^{\mu}\lambda^a q\right)\bigg\rangle\Bigg]   = \pi\alpha_s\times \kappa_{2} \left[ \frac{16}{81}\left(\langle\bar{u}u\rangle^2 + \langle\bar{d}d\rangle^2\right) + \frac{16}{9}\langle\bar{u}u\rangle\langle\bar{d}d\rangle \right]
%\end{multline}
%The first three coefficients $c_i$ $(i=0,1,2)$ are same for the charged and neutral $A_1$ mesons.
Here we study the case of neutral $A_1$ meson only. To simplify the expression for the scalar four-quark condensates factorization technique is adopted.
%\begin{equation}
  %  -\langle(\bar{q_i}\gamma_{\mu}\lambda^a q_j)^2\rangle=\langle(\bar{q_i}\gamma_{\mu}\gamma_{5}\lambda^a q_j)^2\rangle=\delta_{ij}\frac{16}{9}\kappa_j\langle\bar{q_i}q_j\rangle^2
%\end{equation}
The parameter $ \kappa_{1}$ introduces the deviation from the exact factorization which is one in the vacuum saturation assumption \cite{shifman448, zos}. The value of the running coupling constant is $\alpha_s = 0.35$ at $\mu=1$ GeV scale.
For the (neutral) vector meson states $\rho$ and $\omega$, the Wilson coefficients are already given in refs.\cite{91, 100}, which will be used here to study the important effects of (inverse) magnetic catalysis on their masses at finite temperature nuclear matter. In-medium masses of the charged $\rho$ mesons are also calculated to study the in-medium hadronic decay widths of $A_1\rightarrow \rho^{\pm}\pi^{\mp}$, have the corresponding coefficients $c_i; i=0,1,2$, similar to the neutral partner $\rho_0$. However, $c_3$ term is different than that of neutral $\rho$ in the following way,
\begin{equation*}
    c_3^{\rho^{\pm}} = \pi\alpha_s\times \kappa_{2} \left[ \frac{16}{81}\left(\langle\bar{u}u\rangle^2 + \langle\bar{d}d\rangle^2\right) - \frac{16}{9}\langle\bar{u}u\rangle\langle\bar{d}d\rangle \right]
\end{equation*}
%\begin{multline}
 %   c_3^{\rho^{\pm}}=-\frac{\pi}{2}\alpha_s\Bigg[2\left\langle\left(\bar{u}\gamma_{\mu}\gamma^{5}\lambda^a d\right)\left(\bar{d}\gamma_{\mu}\gamma^{5}\lambda^a u\right)\right\rangle   + \frac{2}{9} \bigg\langle\left(\bar{u}\gamma_{\mu}\lambda^a u + \bar{d}\gamma_{\mu}\lambda^ad\right) \times \\
%\left(\sum_{q=u,d,s}\bar{q}\gamma^{\mu}\lambda^a q\right)\bigg\rangle\Bigg]   = \pi\alpha_s\times \kappa_{3} \left[ \frac{16}{81}\left(\langle\bar{u}u\rangle^2 + \langle\bar{d}d\rangle^2\right) - \frac{16}{9}\langle\bar{u}u\rangle\langle\bar{d}d\rangle \right]
%\end{multline}
%The QCD sum rule thus connects the hadronic spectral properties in terms of the resonance parameters given by Eq.(17), to the non-perturbative effects of QCD through the quark and gluon condensates in Eq.(18), with $c_i$'s given by Eqs.(19)-(25).\\
In practice, the parametrization of the current-current correlator is given for the energy region of the lowest hadronic resonance, usually there is no model which can be valid for arbitrary high energies. To achieve a larger suppression on the high energy part of the hadronic spectral distribution, Borel transform is applied.
%It is defined on an arbitrary function $g$ as \cite{k1} 
%\begin{equation}
%    \textit{g}\ (Q^2) \ {\xrightarrow{\hat{B}}} \ \widetilde{\textit{g}}\ (M^2)
%\end{equation}
%\begin{equation}
 %   \hat{B}:= \lim_{Q^2\to\infty, n\to\infty }  \frac{1}{\Gamma(n)} (-Q^2)^n \left(\frac{d}{dQ^2}\right)^n. 
%\end{equation}
%with $Q^2/n= M^2$ become fixed and $M$ is called the Borel mass. %The following relations are useful to determine the Borel transform of Eqs.(17)-(18) 
%\begin{equation}
   % \hat{B}[(Q^2)^{-p}]=\frac{1}{(p-1)!}\frac{1}{(M^2)^p},
%\end{equation}
%\begin{equation}
   % \hat{B}[(Q^2)^{p} ln(Q^2)]=-p!(-M^2)^p,
%\end{equation}
%\begin{equation}
  %  \hat{B}[(Q^2+A)^{-p}]=\frac{1}{(p-1)!}\frac{1}{(M^2)^p}e^{-A/M^2}.
%\end{equation}
After applying the Borel transform, the phenomenological side of Eq.(\ref{pheno}) is connected to the OPE side in Eq.(\ref{ope}) as  
\begin{equation}
   \frac{1}{\pi} \int^{\infty}_0 ds \ e^{-s/M^2} \ Im R^{phen.}(s) =  \left[c_0M^2 + c_1+ \frac{c_2}{M^2} + \frac{c_3}{2M^4} \right]
   \label{boreltransform}
\end{equation}
Where, $M$ is called Borel mass. The exponential function on the l.h.s enhances the contribution of the low energy resonance by suppressing the high energy continuum part at large $s$. Higher dimensional operators in $R_{OPE}$ (on the r.h.s) are suppressed by an additional factor of $1/(n-1)!$, leading to the better convergence of the operator product expansion side.

The hadronic spectral density function $Im R^{phen.} (s)$ separates into a resonance part, $R_{res}(s)$ (for $s\leq s_0$) and a perturbative continuum (for $s> s_0$) \cite{91,leupold,klingl}
\begin{equation}
   \frac{Im R^{phen.} (s)}{\pi}   = R_{res}(s)\ \Theta(s_0 - s) +  c_0 \ \Theta(s- s_0 )
   \label{param}
\end{equation}
Where, $s_0$ is the threshold between the low energy resonance region and the high energy continuum part, latter being calculated in perturbative QCD. 

Next, the finite energy sum rules (FESRs) are derived by inserting Eq.(\ref{param}) into the l.h.s of Eq.(\ref{boreltransform}). The exponential function then expanded in powers of $\frac{s}{M^2}$ for $s\leq s_0$ and $M>\sqrt{s_0}$. Comparing the different powers of $1/M^2$ on both sides of Eq.(\ref{boreltransform}), the finite energy sum rules in vacuum can be written as 
 \begin{equation}
     \int^{s_0}_0 ds \ R_{res}(s) =  (c_0 s_0 + c_1)
 \end{equation}
 \begin{equation}
     \int^{s_0}_0 ds \ s R_{res}(s) =  \left(\frac{c_0 s_0^2}{2} - c_2\right)
 \end{equation}
 \begin{equation}
 \int^{s_0}_0 ds \ s^2  R_{res}(s) =  \left(\frac{c_0 s_0^3}{3} + c_3\right)
 \end{equation}
  The spectral parametrization $R_{A}^{phen.} (s)$, used for the axial-vector current is given by 
  \begin{equation}
           R_{A}^{phen.} (s)= F_{A} \delta(s-m_A^2) + c_0\Theta(s-s_0^A) + f_{\pi}^2\delta(s-m_{\pi}^2)
           \label{a1phen}
     \end{equation}
Where, $\frac{Im R^{phen.} (s)}{\pi}=R^{phen.}(s)$. In the axial-vector meson channel, apart from the $A_1$ resonance term there is a contribution from the pseudoscalar meson resonance, due to its coupling to the axial-vector current \cite{B196, shifman448,leupold}. Similar ansatz is used for the vector meson channel except for the pseudoscalar meson pole \cite{91, klingl, hatsudab394}
\begin{equation}
R_{V}^{phen.}(s) = F_{V} \delta(s-m_V^2) + c_0\Theta(s-s_0^V) 
\end{equation}
%Inserting the expression (33) into Eqs.(30)-(32), the finite energy sum rules for the axial-vector channel, in the vacuum, are given by
%\begin{equation}
 %   F_A= \left(c_0 s_0^A + c_1 - f_{\pi}^2\right)
  %  \label{fsras}
%\end{equation}    
%\begin{equation}
 %   F_Am_A^2= \left(\frac{c_0 (s^A_0)^2}{2} - c_2 - f_{\pi}^2m_{\pi}^2\right)
%\end{equation}
%\begin{equation}
 %   F_A m_A^4 = \left(\frac{c_0 (s_0^A)^3}{3} + c_3^A -f_{\pi}^2m_{\pi}^4\right)
 %   \label{fsrae}
%\end{equation}
%In the same way, for the vector meson channel in the vacuum, using Eq.(34) into Eqs.(30)-(32), one obtains 
%\begin{equation}
 %   F_V= \left(c_0 s_0^V + c_1 \right)
 %   \label{fsrvs}
%\end{equation}    
%\begin{equation}
%    F_Vm_V^2= \left(\frac{c_0 (s^V_0)^2}{2} - c_2 \right)
%\end{equation}
%\begin{equation}
 %   F_V m_V^4 = \left(\frac{c_0 (s_0^V)^3}{3} + c_3^V \right)
  %  \label{fsrve}
%\end{equation}

In the nuclear medium, for mesons at rest, the meson-nucleon scattering effect is incorporated through the Landau damping term $\rho_{sc}$ in the spectral function as given by,

\begin{equation}
  \int^{\infty}_0 ds \ e^{-s/M^2} \ \frac{Im R^{phen.}(s)}{\pi}  + \rho_{sc} =  \left[c_0M^2 + c_1+ \frac{c'_2}{M^2} + \frac{c'_3}{2M^4} \right]
  \label{fesrfinal}
\end{equation}
Where the primed symbols denote the corresponding in-medium quantities. The damping term originates due to the absorption of a space-like meson by an on-shell nucleon \cite{leupold, flower}. The contributions for the vector and axial-vector meson channels are taken according to the formalism based on the scattering amplitudes with the nucleons of the respective polarization states of vector and axial-vector correlators \cite{flower}. 
%Incorporating the scattering term as, $\rho_{sc}=\frac{\rho_B}{4M_N}$ \cite{91, klingl, prc52, flower} in two channels (for both charged and neutral mesons with different $c'_3$), the modified FESRs for the axial-vector channel in medium  are 
%\begin{equation}
 %   F'_A= \left(c_0 s_0^{'A} + c_1 - f_{\pi}^2 - \frac{\rho_B}{4M_N}\right)
 %   \label{fesrs}
%\end{equation}    
%\begin{equation}
 %   F'_Am_A^{'2}= \left(\frac{c_0 (s^{'A}_0)^2}{2} - c'_2 - f_{\pi}^2m_{\pi}^2\right)
%\end{equation}
%\begin{equation}
 %   F'_A m_{A}^{'4} = \left(\frac{c_0 (s_0^{'A})^3}{3} + c_3^{'A} -f_{\pi}^2m_{\pi}^4\right)
%\end{equation}
%The in-medium FESRs for the vector meson channel \cite{91} are thus modified to
%\begin{equation}
 %   F'_V= \left(c_0 s_0^{'V} + c_1 - \frac{\rho_B}{4M_N}\right)
%\end{equation}    
%\begin{equation}
 %   F'_Vm_V^{'2}= \left(\frac{c_0 (s^{'V}_0)^2}{2} - c'_2 \right)
%\end{equation}
%\begin{equation}
 %   F'_V m_V^{'4} = \left(\frac{c_0 (s_0^{'V})^3}{3} + c_3^{'V} \right)
  %  \label{fesre}
%\end{equation}
The coefficient in the scalar four-quark condensate $\kappa_j$ for the vector and axial-vector states, are determined by solving their respective vacuum FESRs. Thus, the in-medium resonance parameters of the vector and axial-vector spectral functions, i.e., their corresponding mass $m'$, strength $F'$ and threshold energy $s'_0$, are solved from the in-medium FESRs as derived from Eq.(\ref{fesrfinal}). 

\section{In-medium Hadronic Decay widths}
\label{sec4}
An effective Lagrangian involving the axial-vector (a), vector (v) and pseudoscalar ($\phi$) mesons vertices, is used in the present investigation to calculate the in-medium partial decay width of $A_1$ meson going to $\rho$ and $\pi$ mesons states. The effects of (inverse) magnetic catalysis on the in-medium hadronic decay widths of $A_1\rightarrow \rho \pi$ are obtained through the in-medium mass of the initial state $A_1$ and final state $\rho$ mesons particles, calculated using QCD sum rule approach. The effects of temperature in the magnetized nuclear matter is considered in the present study, including the effects of magnetized Dirac sea contribution. The phenomenological Lagrangian used for the $av\phi$ interaction 
\begin{equation}
      \mathcal L_{{av\phi}}= i \tilde{f}  \langle a_{\mu\nu}[v^{\mu\nu}, \phi]\rangle.
      \label{avp}
 \end{equation}
Where, the $SU(3)$ multiplet of the mesons are considered; The symbol $<>$ denotes the trace of the  product of matrices and $i$ in front of $\tilde{f}$ is to make the Lagrangian hermitian. The spin-1 meson fields in Eq.(\ref{avp}) are treated as anti-symmetric tensor fields.
% $a_{\mu\nu}$ and $v_{\mu\nu}$ which transform under a non-linear realization of chiral symmetry group $G=SU(3)_L\times SU(3)_R $ as \cite{eckerb321},
%\begin{equation}
%    X_{\mu\nu} \ {\xrightarrow{G}}\ h(\phi') X_{\mu\nu} h(\phi')^{\dagger}, \quad X_{\mu\nu}=a_{\mu\nu}, v_{\mu\nu}
%\end{equation}
%The chiral $SU(3)_L\times SU(3)_R $ symmetry group is spontaneously broken down to $SU(3)_V$. A non-linear realization of $G$ is defined on the elements $u(\phi')$ of the coset space $SU(3)_L\times SU(3)_R/ SU(3)_V $ as \cite{coleman}
%\begin{equation}
%    u(\phi') \ {\xrightarrow{G}}\ g_R u(\phi') h(\phi')^{\dagger}=h(\phi') u(\phi') g_L^{\dagger};\ \ g_{R, L}\in SU(3)_{R,L},\  h(\phi')\in SU(3)_V
%\end{equation}
%where $\phi'$ are the Goldstone boson fields and $\phi=\frac{1}{\sqrt{2}}\sum_{i=1}^8\lambda_i\phi_i'$. The elements of the coset space are defined as, $u(\phi')=exp(-\frac{i}{\sqrt{2}}\frac{\phi}{f})$. The vector, axial-vector and pseudoscalar meson fields transform as octets under $SU(3)_V$. If the $SU(3)$ multiplets are denoted by $S$ then, the non-linear realization of $G$ leads to the following transformations 
%\begin{equation}
%   S \ {\xrightarrow{G}}\ h(\phi') S h(\phi')^{\dagger}, \quad S= a, v, \phi 
%\end{equation}
%The transformations, Eqs.(49)-(51) preserve the symmetry of the Lagrangian. Thus, a phenomenological Lagrangian is used for the $A_1\rho\pi$ vertices (along with the other vertices in the octet of mesons), to find the decay widths of $A_1\rightarrow \rho\pi$ channels \cite{roca70}.
The channel of neutral axial-vector meson $A_1^0 \rightarrow \rho^{\pm}\pi^{\mp}$ is studied in the hot magnetized matter, accounting for the effects of (inverse) magnetic catalysis. The amplitude for the process is calculated using the above Lagrangian.

%To calculate the hadronic decay widths for various $(A_1\rightarrow \rho \pi)$ channels, the amplitudes are derived from the above Lagrangian. 

%Among the various possible strong decay modes of axial-vector mesons in the meson nonet, decaying into vector and pseudoscalar mesons, some are only observed, no exact data is available for those decay widths \cite{pdg, roca70}. 
The coupling parameters in the interaction vertices are fitted from the observed decay modes of $J^{PC}=1^{++}$ and $J^{PC}=1^{+-}$ family of axial-vector mesons with their respective branching ratio \cite{roca70, pdg}. %These fittings are described in detail in that reference using both $J^{PC}=1^{+-}$ and $J^{PC}=1^{++}$ multiplets of axial-vector mesons, namely the class of $B_1$ and $A_1$ mesons. 
The value of the fitted parameter $\tilde{f}$ in Ref. \cite{roca70}, is used here to find the in-medium partial decay widths of the $A_1^0$ meson in an external medium. For tree level calculations $(\partial_{\mu}X_{\nu}-\partial_{\nu}X_{\mu})$ can be used instead of the tensor $X_{\mu\nu}$ of $X = a, v$ fields. Normalization leads to \cite{eckerb321}
 \begin{equation}
     \langle 0 | X_{\mu\nu} | X; p, \epsilon \rangle = \frac{i}{m_X}[p_{\mu}\epsilon_{\nu}(X)-p_{\nu}\epsilon_{\mu}(X)].
     \label{norm}
 \end{equation}
  $m_X$ is the mass of the $X$ meson. The decay width of $A_1 \rightarrow \rho\pi$ channel is given by
 \begin{equation}
     \Gamma_{A_1 \rightarrow \rho\pi} = \frac{q}{8\pi m_{A_1}^2} \overline{|\mathcal{M}|^2} 
     \label{decay}
 \end{equation}
 Where, $q$ is the momentum of the final state particles in the rest frame of $A_1$ meson 
 \begin{equation}
     q(m_{A_1}, m_{\rho}, m_{\pi}) = \frac{1}{2 m_{A_1}} \Big([m_{A_1}^2-(m_{\rho}+ m_{\pi})^2] \times [m_{A_1}^2-(m_{\rho}- m_{\pi})^2] \Big)^{1/2}
     \label{momentum}
 \end{equation}
 The interaction terms corresponding to the $A_1\rho\pi$ vertices as given by Lagrangian (\ref{avp}), are
 \begin{equation}
    \mathcal{L}_{A_1\rho\pi} = \sqrt{2}i\tilde{f} \Big[ A_1^0\left(\rho^-\pi^+ -\rho^+\pi^-\right) + A_1^+\left(\rho^+\pi^0 - \rho^0\pi^+ \right) + A_1^-\left(\rho^0\pi^- -\rho^-\pi^0 \right) \Big] 
    \label{lagrangian}
 \end{equation}
 These terms result from the parity and charge conjugation symmetry of the Lagrangian density.
 The amplitude obtained from the interaction vertices is thus given as 
 \begin{equation}
     \mathcal{M} = \frac{-2\lambda_{av\phi}}{m_{A_1}m_{\rho}} \left(p'.p \quad\epsilon'.\epsilon - \epsilon'.p\quad \epsilon.p'\right)
     \label{amp}
 \end{equation}
 With $p'$, $ \epsilon'$ and $p$, $\epsilon$ are the four-momenta and polarization vectors of the $A_1$ and $\rho$ mesons, respectively, and $\lambda_{av\phi}=\sqrt{2}i \tilde{f}$. Using the normalization of the fields [eq.(\ref{norm})], the gauge invariant amplitude is obtained.  %Therefore, the decay width is given by eq.(\ref{decay}) by using eqs.((\ref{momentum})-(\ref{amp})) and the invariant relations between four-momenta, $p^2=m_{\rho}^2$, $p'^2=m_{A_1}^2$ and $2p.p'=(m_{A_1}^2+m_{\rho}^2-m_{\pi}^2)$,
 The decay width of $A_1\rightarrow \rho \pi$ is therefore written as
 \begin{equation}
     \Gamma_{A_1 \rightarrow \rho \pi} =  \frac{|\lambda_{av\phi}|^2}{2\pi m_{A_1}^2}q\left(1+\frac{2q^2}{3m_{\rho}^2}\right)
     \label{finaldecay}
 \end{equation}
 To consider the decay of the neutral light vector meson $\rho^0$ into two pseudoscalar mesons $\pi^{+}\pi^-$, an effective vector-pseudoscalar interaction Lagrangian is taken into account with the tree-level coupling constant $\tilde{g}$ fitted experimentally by the observed decay width of $\rho\rightarrow \pi^+\pi^-$. The tree level amplitude leads to the following decay width \cite{klingl2}
 \begin{equation}
 \Gamma_{\rho\rightarrow \pi^+\pi^-}=\frac{\tilde{g}^2}{48\pi}m_{\rho}\left(1-\frac{4m_{\pi}^2}{m_{\rho}^2}\right)^{3/2}
     \label{rhodecay}
 \end{equation}

 \section{Spectral Function and Production Cross-sections}
 \label{sec5}
 
 The normalized form of the relativistic Breit-Wigner spectral function for the light vector and axial-vector mesons are considered as \cite{amspec, elena}
 \begin{equation}
     A(M) = \frac{2}{\pi}\frac{M^2\Gamma^*}{(M^2 - m^{*2})^2 - M^2\Gamma^{*2}}.
     \label{spec}
 \end{equation}
 Where, $M$ is the invariant mass $m^*$ and $\Gamma^*$ are the in-medium mass and decay widths of the corresponding vector and axial-vector meson states. The relativistic Breit-Wigner cross-section for productions of neutral $\rho$ meson from the scattering of $\pi^+$ and $\pi^-$ mesons and $A_1$ in the $\rho^+\pi^-$ and $\rho^-\pi^+$ channels, in a hot magnetized nuclear matter is given by
 \begin{equation}
     \sigma (M) = \frac{6\pi^2\Gamma^*A(M)}{q(m^*, m_1, m_2)^2}.
     \label{cross}
 \end{equation}
 Here, $q$ is the momentum of the scattering particles in the center-of-mass frame of the produced particle \cite{amspec}. In eq.(\ref{cross}), $m_{1}$, $m_{2}$ are the masses of the scattering particles. The spectral functions and production cross-sections of $\rho^0$ and $A_1^0$ mesons are studied in the present work by incorporating the effects of the magnetized Dirac sea to the in-medium masses as described in sections (\ref{sec2}) and (\ref{sec3}). This effect leads to the phenomena of (inverse) magnetic catalysis at vacuum and at nuclear matter saturation density $\rho_0$, depending on the finite or zero values of the anomalous magnetic moments of the nucleons. The effects are studied both at zero and finite temperature nuclear matter in a background magnetic field for baryon density $\rho_B=0$ and $\rho_0$. As the density of the medium in a strong magnetic field in the peripheral ultra-relativistic heavy ion collision is typically very low. In-medium masses are calculated within the QCD sum rule approach by incorporating the medium effects through the light quark (up to the scalar four quark) condensates and the scalar gluon condensates, which are obtained from the chiral effective model. In-medium decay widths of the processes $A_1^0 \rightarrow \rho^{\pm}\pi^{\mp}$ and $\rho^0 \rightarrow \pi^+\pi^-$ are computed using the appropriate form of the effective interaction Lagrangian involving the parent and daughter particles as illustrated in section (\ref{sec4}).

 \section{Results and Discussions}
 \label{sec6}
In the present work, the in-medium spectral properties of mass and decay widths of the light vector mesons $\omega$, $\rho$ and the light axial-vector meson $A_1$ are investigated in the asymmetric nuclear medium at finite temperature, in presence of an external magnetic field incorporating the effects of the magnetized Dirac sea. Using the in-medium mass and decay widths of the mesons, the spectral functions as well as the production cross-sections of the neutral $\rho$ and $A_1$ mesons are studied from their respective scattering modes (also decay modes) of $\rho \rightarrow \pi^{+}\pi^-$ and $A_1\rightarrow \rho^{\pm}\pi^{\mp}$. The masses are computed using the QCD sum rule (QCDSR) approach in terms of the in-medium values of the light quark condensates, the scalar gluon condensates, and the parameters of the QCD Lagrangian, namely the current quark masses $m_u$, $m_d$, the QCD running coupling constant $\alpha_s$. In-medium values of the QCD condensates are obtained from that of the scalar isoscalar fields $-$ non-strange $\sigma$, strange $\zeta$, the scalar isovector field $\delta$ and the scalar dilaton field $\chi$, within the chiral $SU(3)$ model [as given by eqs.(\ref{uu})-(\ref{gg})]. In the chiral $SU(3)_L\times SU(3)_R$ model, the meson fields are treated to be classical and the nucleons as the quantum fields in obtaining the one-loop self energy functions of the nucleons with the magnetized fermionic propagator in the weak-field limit. In this way the effects of the Dirac sea at finite magnetic field are taken into account. The effects of temperature are incorporated through the Fermi distribution functions in the number and scalar densities of protons and neutrons in magnetized matter, eqs.(\ref{rhp})-(\ref{rhsn}). The coupled equations of motion of the scalar fields are then solved at the given values of baryon density $\rho_B$, isospin asymmetry $\eta  \big(=\frac{\rho_n - \rho_p}{2\rho_B}\ \big)$, temperature $T$ and magnetic field $|eB|$, through the number ($\rho_{p,n}$) and scalar densities ($\rho^s_{p,n}$) of protons and neutrons. Furthermore, the energy levels of the charged nucleons, i.e., protons are modified by the magnetic field, giving rise to the Landau energy levels. There is another effect coming from the magnetic field due to the anomalous magnetic moments (AMMs) of the nucleons in both the Fermi and Dirac sea of nucleons. The contribution of an external magnetic field on the Dirac sea are obtained through summation over the nucleonic tadpole diagrams due to their interactions with the scalar meson fields $\sigma$, $\zeta$ and $\delta$ in the chiral effective model Lagrangian. The fermionic propagator are taken up to second order expansion in magnetic field $|eB|$ in the calculation of the one loop self energy functions of the nucleons, along with the non-zero anomalous magnetic moments of the nucleons. Thus, the Dirac sea contributes to the scalar densities of the nucleons. In ref.\cite{chrmmc}, the effects of magnetized Dirac sea have been studied on the in-medium mass of the lowest lying states of heavy quarkonia at finite density and zero temperature. In the present investigation, we have studied the effects of the magnetized Dirac sea on the in-medium spectral properties of the light vector and axial-vector mesons in asymmetric nuclear matter at finite temperature.

 The light quark condensates ($\langle \bar{q}q \rangle$; $q=u,d$) and the scalar gluon condensate $(\langle\frac{\alpha_{s}}{\pi}G_{\mu\nu}^a G^{a\mu\nu}\rangle)$, thus obtained are used to calculate the in-medium mass and other spectral parameters of the light axial-vector and vector mesons, by solving their respective finite energy sum rules. 
 %[eqs.{(\ref{fesrs})-(\ref{fesre})}].  
 %The values of the current quark masses, running coupling constant, to be used in the present study are $m_u=4$ MeV, $m_d=7$ MeV and $\alpha_s=0.3551$ \cite{91, 100}. 
 The vacuum masses of the light mesons are taken as per the particle data group values of $m_{\rho}=770$ MeV, $m_{\omega}=783$ MeV and $m_{A_1}=1230$ MeV, \cite{pdg}. Using these values the vacuum finite energy sum rules %(eqs.{(\ref{fsras})-(\ref{fsrae})} 
 for $\rho$, $\omega$ and $A_1$ %eqs.{(\ref{fsrvs})-(\ref{fsrve})} for 
 are solved to obtain the respective coefficients $\kappa_j$ of the scalar four quark condensates. The values obtained are: $\kappa_1=-2.204$ for $A_1^0$; %$\kappa_2=-2.485$ (for $A_1^{\pm}$); and 
 $\kappa_2=-8.804$ for $\rho^{\pm}$, $\kappa=7.237$ and $\kappa=7.789$ for $\rho^0$ and $\omega$ mesons respectively. In the QCD sum rule study of the light vector mesons at strange hadronic matter \cite{91}, the values of $\kappa_j=7.236, 7.788, -1.21$ were obtained for $j=\rho^0,\omega, \phi$ mesons respectively, by solving their corresponding vacuum FESRs. For the case of $A_1$ meson there is an extra pion pole contribution in its spectral density function eq.(\ref{a1phen}). The pion decay constant is $f_{\pi}=93.3$ MeV and the pion mass $m_{\pi}=139$ MeV, remain fixed as in the context of chiral effective model Lagrangian \cite{papa}. The vacuum values of the perturbative continuum threshold $s_0$ and the resonance strength $F$ (both in $GeV^2$) are thus found to be $1.266$ and $2.114$ (for $\rho^0$ state), $2.268$ and $2.755$ (for $A_1^0$ state) and $1.3046$ and $0.2419$ (for $\omega$ state), respectively. The nuclear matter saturation density of $\rho_0=0.15\ fm^{-3}$ is used in the present study. 
 \begin{figure}
    \centering
\includegraphics[width=1.0\textwidth]{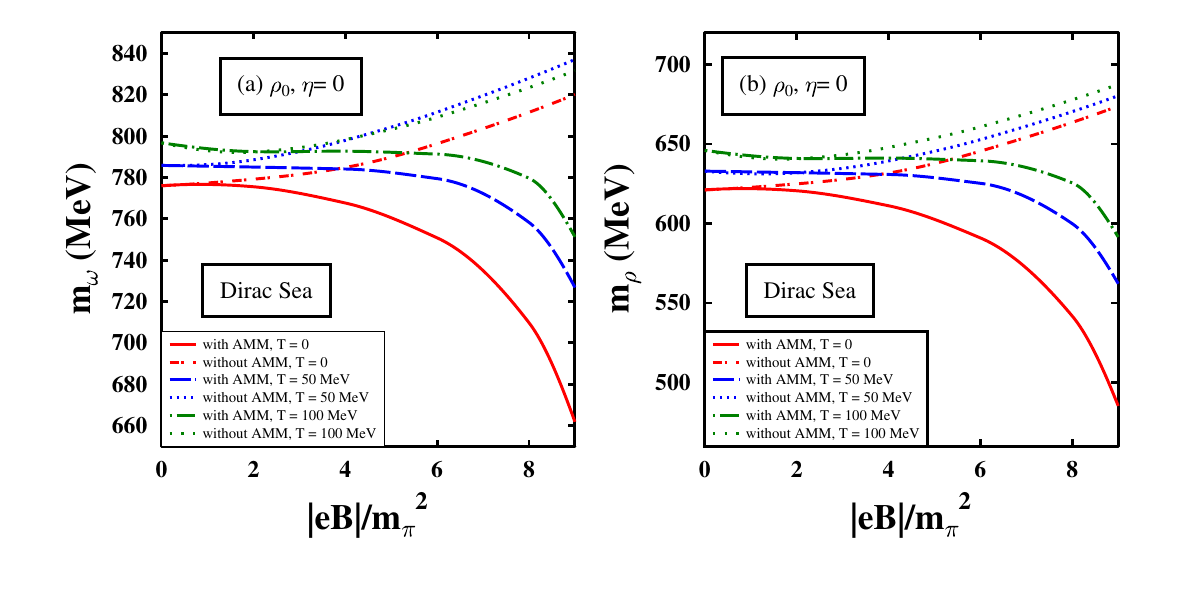}
    \vspace{-1.5cm}
    \caption{\raggedright{ In-medium masses of the $\omega$ and $\rho^0$ mesons are plotted as functions of $|eB|/m_{\pi}^2$ at $\rho_0$ and $\eta=0$ for $T=0,\ 50,\ 100$ MeV. The effects of the magnetized Dirac sea are shown on the masses. Effects of the nucleons' AMMs are considered and compared to the case with zero anomalous magnetic moment.}} 
    \label{1}
\end{figure}
\begin{figure}
    \centering
\includegraphics[width=1.0\textwidth]{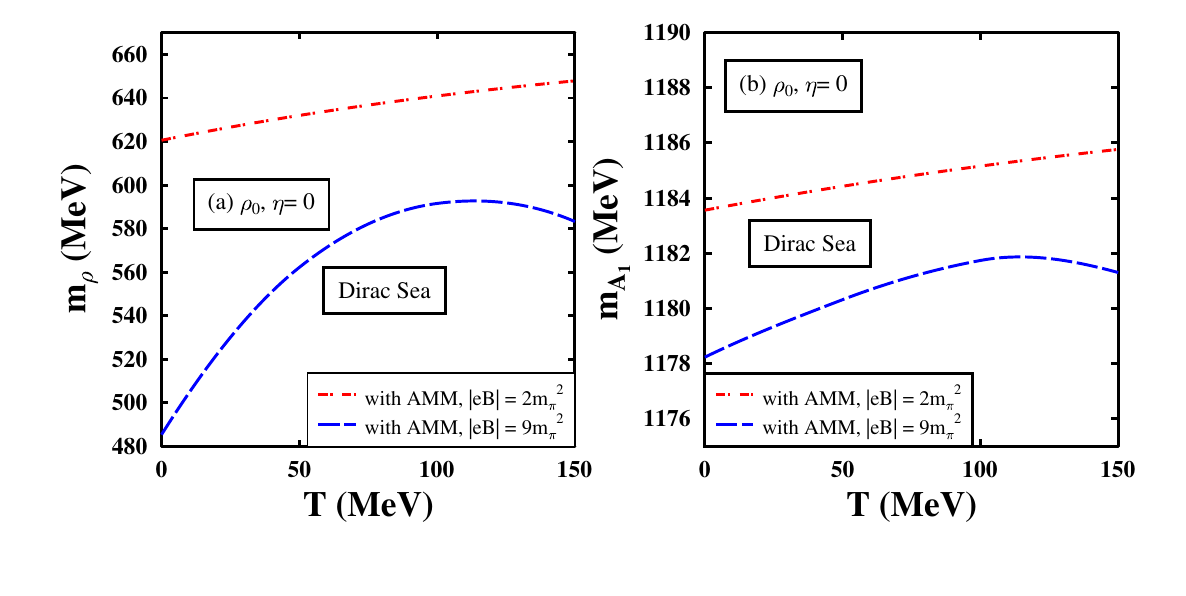}
    \vspace{-1.5cm}
    \caption{\raggedright{ In-medium masses of the $\rho$ and $A_1$ mesons are plotted as functions of temperature T (in MeV) for $|eB|/m_{\pi}^2=2,\ 9$ at $\rho_0$ and $\eta=0$. The effects of the magnetized Dirac sea are incorporated with non-zero nucleons' AMMs.}} 
    \label{1.1}
\end{figure}

 Fig.\ref{1} illustrates the in-medium masses of the (a) $\omega$ and (b) $\rho$ mesons as functions of magnetic field $|eB|$ (in units of $m_{\pi}^2$), in the symmetric nuclear matter ($\eta=0$) at $\rho_0$ for zero as well as finite temperature $T=0,\ 50,\ 100 $ MeV. The anomalous magnetic moments of the nucleons are considered in the magnetized Dirac sea effects and comparison is shown to the case when the AMMs are taken to be zero. The behavior obtained can be explained from the underlying trend of variation of the light quark and the scalar gluon condensates with changing magnetic field at different medium conditions. At $\rho_0$, for zero as well as finite temperatures the values of the light quark condensates, the scalar gluon condensate decrease with increasing magnetic field for non zero AMMs of the nucleons, an effect called inverse magnetic catalysis. The opposite behavior is obtained when nucleonic AMM is taken to be zero, lead to magnetic catalysis. This is observed for both symmetric and asymmetric nuclear matter, for e.g., for pure neutron rich matter with $\eta=0.5$. At $\rho_B=0$, there is no additional contribution from the protons Landau energy levels, any effect due to the magnetic field is obtained through the magnetized Dirac sea contribution only. At zero density, the light quark condensates increase with magnetic field for both finite and zero anomalous magnetic moments of the Dirac sea of nucleons, leading to the effect known as magnetic catalysis. There is observed to be considerable amount of difference in the behavior of the light quark condensates for finite values of the AMMs of the Dirac sea of nucleons as compared to the zero AMM case.
 
 In fig.\ref{1.1}, the mass of (a) $\rho$ and (b) $A_1$ mesons are plotted as functions of temperature (in MeV) at the nuclear matter saturation density $\rho_0$, $\eta=0$. The masses of both $\rho$, $A_1$ mesons tend to rise with temperature by very small amount at the magnetic field strength of $2m_{\pi}^2$. However, at the field strength of $9m_{\pi}^2$ (plotted as long dashed line), the masses initially increase with temperature with a sudden drop after 100 MeV till 150 MeV. The initial rise in mass with temperature may be due to the contribution from the higher momenta region in the three-momentum integral of $\rho_{p,n}$ and $\rho^s_{p,n}$, within the context of chiral $SU(3)$ model at finite density and temperature. However, at high enough temperatures the condensates tend to decrease as it approach to the critical temperature of chiral symmetry restoration. In fig.\ref{1.1} the effects of inverse magnetic catalysis on the masses are observed at finite temperature and density, as it can be inferred from the fact that the overall mass decreases from $2m_{\pi}^2$ to $9m_{\pi}^2$, at any fixed temperature. At high magnetic fields, the condensates decrease at high temperature region towards the chiral symmetry restoration, an effect induced by the phenomenon of inverse magnetic catalysis in hot magnetized nuclear matter. %The scalar fields solutions at $\rho_B=0$ and for nonzero AMMs, are obtained up to $|eB|=3.9\ m_{\pi}^2$ in our present study. 
 
 In the present work, we have investigated the effects of the magnetized Dirac sea at finite (but low) density and temperature (below critical temperature of chiral transition) medium over a range of magnetic field from $|eB|= 0$ to $9 m_{\pi}^2$. The variation is shown in figs.\ref{1}-\ref{1.1} until the behavior of the condensates starts fluctuating with respect to the change in magnetic field. As we have applied a weak-field expansion of the nucleonic propagator up to the second order in magnetic field, it gives reliable results till $|eB| = 3.9 m_{\pi}^2$ at $\rho_B = 0$, and $|eB| = 9 m_{\pi}^2$ at $\rho_B = \rho_0$, $\eta= 0, 0.5$ for non-vanishing AMMs of the Dirac sea of nucleons at $T=0$ to 150 MeV. Now, this range gives considerably smooth variation of the condensates with magnetic field, leading to the phenomena of magnetic (inverse) catalysis at the specific conditions of baryon density and AMMs of the Dirac sea of nucleons. Considerably the region of valid field strength is consistent with that of the hadronic freeze-out hypersurface (FOHS), where the light mesons e.g., pseudoscalar $\pi$, vector $\rho$ mesons etc., are profusely produced. At FOHS, the estimated magnetic field is as low as $4 m_{\pi}^2$ \cite{fohs}, thus the weak-field expansion is well applicable in this regime. In ref.\cite{fohs}, it has been studied that the impacts of magnetic field on the charged light mesons e.g., $\pi^{\pm}$, $\rho^{\pm}$, should have significant observable consequences at the freeze-out hypersurface. The amount of particle production and particle ratio at FOHS can be used to estimate the field strength, lifetime (still a major open question to search for) of the strong and transient magnetic field which is expected to be produced at the very early stages of the non-central heavy ion collision experiments at RHIC, LHC. Thus, the evolution and magnitude of the produced magnetic field at FOHS can be estimated from the ratio of the production number of charged particles (e.g., $\rho^{\pm}/\pi^{\pm}$). As their production are strongly dependent on their masses having strong magnetic field dependencies, this may lead to their modified yields at freeze-out. The study of magnetic field effects of (inverse) magnetic catalysis on the light vector and axial-vector mesons, should have considerable amount of observable impacts, e.g., in the production of different light mesons species at the stage of freeze-out in the non-central heavy ion collision experiments. However, at finite density and finite temperature medium, the validity range is somewhat extended covering the region at the freeze-out hypersurface.  

Hadron mass e.g., that of the mesons is, in general, not a directly experimentally observable quantity. However, from the in-medium mass and decay widths the spectral function can be constructed. In general, the spectral function then receives the peak shift, width broadening, appearance of new peaks etc., due to the various in-medium effects in the relativistic heavy ion collisions. The in-medium masses of the $\omega$, $\rho$ and $A_1$ mesons at different values of temperature and magnetic fields, as shown in figs.\ref{1}-\ref{1.1}, give an impact of the effects of magnetic (inverse) catalysis on the in-medium properties of light hadrons. The masses are calculated using the QCD sum rule approach as described in section \ref{sec3}, in terms of the in-medium light quark condensates (up to scalar four quark) and the scalar gluon condensate. The light quark condensates increase (decrease) with magnetic field due to the phenomena of magnetic (inverse) catalysis, which is observed to be an important effect in the present analysis of light meson properties in a background magnetic field.

In the chiral $SU(3)$ model the critical temperature of chiral phase transition or a crossover is obtained around $T_c=150$ MeV. This phase transition is of purely hadronic nature as the model involves only hadronic degrees of freedom. %Although, the characteristics e.g., order, latent heat etc., of different phase transitions depend on the particular set of parameters fitted from the empirically known quantities e.g., baryon mass etc. %Therefore, the pseudo-critical temperature of chiral symmetry restoration can be considered as $150$ MeV within the chiral $SU(3)$ model. This is estimated to be around $150\pm 20$ MeV from lattice QCD study, and $155$ MeV (at the vanishing chemical potential) from some theoretical calculations.
In ref.\cite{borsanyi}, the transition temperature is found to be around $T_c=155$ MeV %[the statistical and systematic errors are not shown here]
from the inflection point of chiral condensate using the lattice QCD simulation at zero baryon chemical potential $\mu_B=0$. %The physical quark masses have been used in the simulation with lattice resolutions ranging from $N_t=6$ to 16 [$N_t$ denotes the number of lattice points in the Euclidean time direction, used to find the lattice spacing]. In this case the transition is of a crossover type, the associated critical temperature depends upon the choice of the observables. E.g., the chiral condensate act as an order parameter at the vanishing quark masses to determine the chiral transition temperature. 
From an estimate of the chiral condensate at the pseudo-critical temperature it has been observed that its value is about half of that in vacuum, which is also close to the one estimated at $2\rho_0$ in the cold nuclear matter \cite{holt}. In the presence of a magnetic field background at $\mu_B=0$, the magnetic field variation of the chiral condensates can be obtained from the first principle calculations of lattice QCD simulations. The lattice results indicate that the critical temperature increases monotonically with increasing magnetic field. %The confinement theory of QCD at low energies are described using various effective theory models based on the underlying symmetry principles of strong interaction physics.
%The light quark condensate is usually related to the non-vanishing expectation value of the scalar isoscalar field $\sigma$ in chiral effective model Lagrangian, represented by a composite operator of quark-antiquark fields.
The enhancement of the QCD light quark condensates with the magnetic field is named as magnetic catalysis (MC), while its opposite behavior is known as inverse magnetic catalysis (IMC). %although most of the model calculations show MC there are some lattice calculations which support the opposite behavior of inverse magnetic catalysis (IMC). 
It has been illustrated that the effect of IMC is attributed to the dominance of the sea contribution over the valence term of the quark condensate, e.g., in the context of Nambu-Jona-Lasinio (NJL) model, quark-meson models, etc. The values of the light quark condensates tend to decrease with increasing temperature and/or density of the surrounding medium. Consequently, the enhancement in the critical temperature corresponds to the increment in the quark condensate. However, some lattice results show IMC at specific medium conditions, correspondingly the critical temperature shows opposite behavior compared to the case in magnetic catalysis. In our present study, at zero baryon density $\rho_B=0$, effects of the magnetized Dirac sea lead to the phenomenon of magnetic catalysis for both non-zero AMM (till $3.9m_{\pi}^2$) and zero AMM of the Dirac sea of nucleons below critical temperature. At $\rho_B=\rho_0$, inverse magnetic catalysis is observed for non-vanishing nucleonic AMMs, which turn out to be magnetic catalysis for zero AMM, both at zero and finite temperatures. Notably, the current study concerns with the behavior of the light vector and axial-vector mesons in a magnetized hot nuclear matter, accounting for the effects of (inverse) magnetic catalysis [coming from magnetized Dirac sea] below the pseudo-critical temperature of chiral phase transition.

The in-medium partial decay width of $A_1\rightarrow\rho\pi$ process is calculated from eq.(\ref{finaldecay}), by using the experimentally fitted parameters in the Lagrangian, masses of the initial and final states particles. The mass of the neutral and charged pions to be used here are $134.97$ MeV and $139.57$ MeV respectively \cite{pdg}. The vacuum decay widths for $A_1^0\rightarrow\rho^{\pm}\pi^{\mp}$ channels are calculated to be $202.58$ MeV for $\tilde{f}= 1540$ MeV, obtained from experimentally fitted values of the relevant decay modes as mentioned in table.(4) of ref.\cite{roca70}. The value is very close to the central average value of $210$ MeV (corresponding to the $50\%$ branching ratio) for $A_1\rightarrow\rho\pi$ channel. The phenomenological interaction Lagrangian contains no term involving $A_1^{0}\rho^0\pi^0$ vertex, thus there is no decay for $A_1^0\rightarrow \rho^0\pi^0$ mode. The authors of CLEO Collaboration \cite{tao} have been reported a branching ratio of about $60\%$ for the intermediate $\rho\pi$  ($S$-wave) state of $A_1^{-}$ meson decaying into $\pi^- \pi^0 \pi^0$ relative to the total $A_1^-\rightarrow \pi^- \pi^0 \pi^0$ decay. The works of \cite{chen91} showed $\rho\pi$ ($S$-wave) to be the dominant decay channel for the ground state of $A_1$ meson in its class of axial-vector mesons with quantum numbers $(I^G(J^{PC})=1^-(1^{++}))$. They have analyzed the mass spectra of all the possible light axial-vector meson states by Regge trajectory analysis and have applied a light quark pair creation model \cite{micu} to calculate their OZI-allowed strong decays. 
 
\begin{figure}%
    \centering
    \subfloat[\centering ]{{\includegraphics[width=8.4cm]{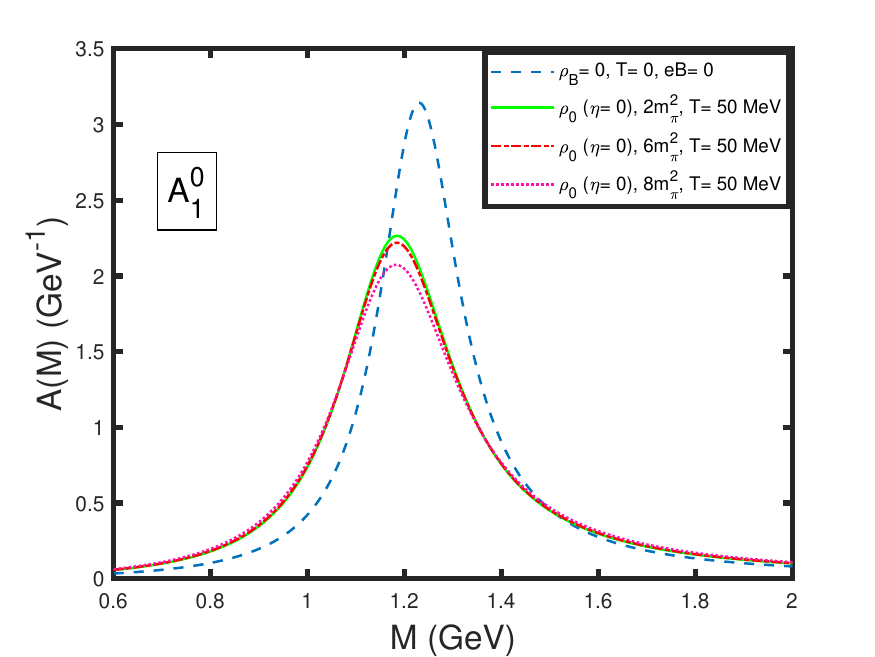} }}%
    \hspace{-0.8cm}
    \subfloat[\centering ]{{\includegraphics[width=8.4cm]{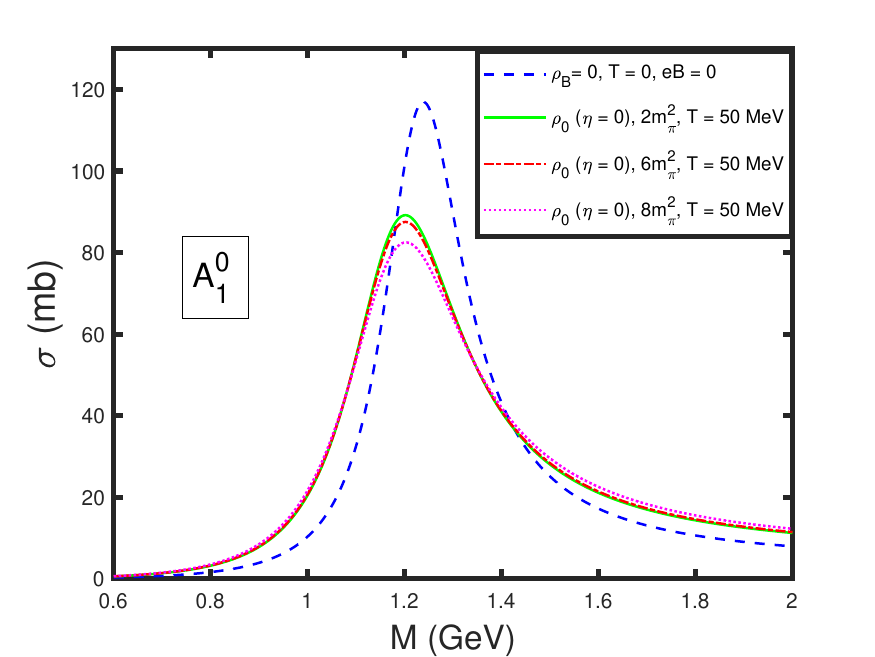} }}%
    \vspace{-0.4cm}
    \caption{\raggedright{ (a) Spectral function $A(M)$ (in Ge$V^{-1}$), (b) production cross-section $\sigma$ (in mb) of the neutral $A_1$ meson are plotted as functions of the invariant mass $M$ (in GeV) at $\rho_B=\rho_0$, in the symmetric nuclear matter ($\eta=0$), for $|eB|/m_{\pi}^2=2,6,8$ at $T=50$ MeV. The vacuum case with $\rho_B=0$, $T=0$ also $|eB|=0$ is shown for comparison as the dashed line. }}%
    \label{2}%
\end{figure}

\begin{figure}%
    \centering
    \subfloat[\centering ]{{\includegraphics[width=8.4cm]{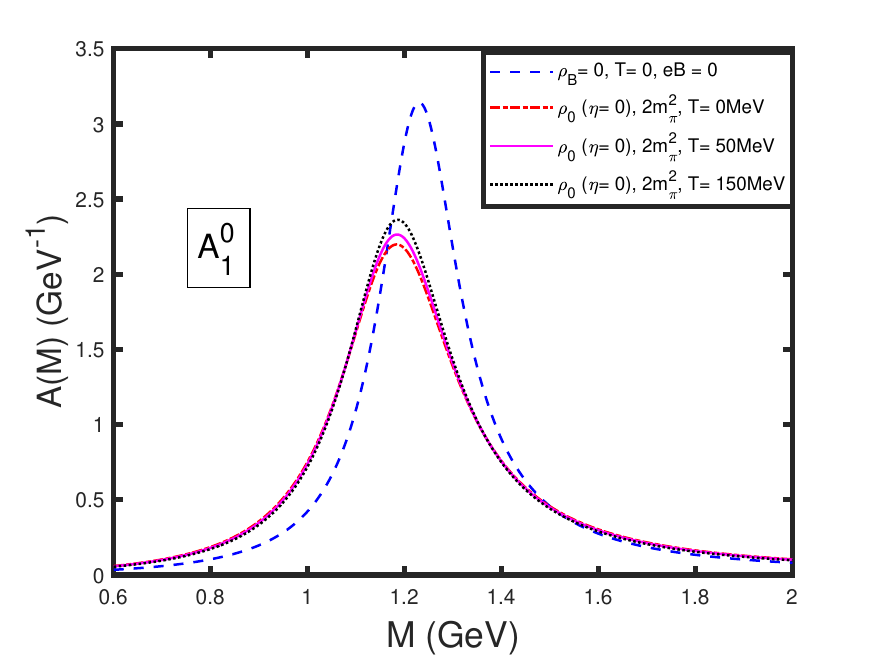} }}%
    \hspace{-0.8cm}
    \subfloat[\centering ]{{\includegraphics[width=8.4cm]{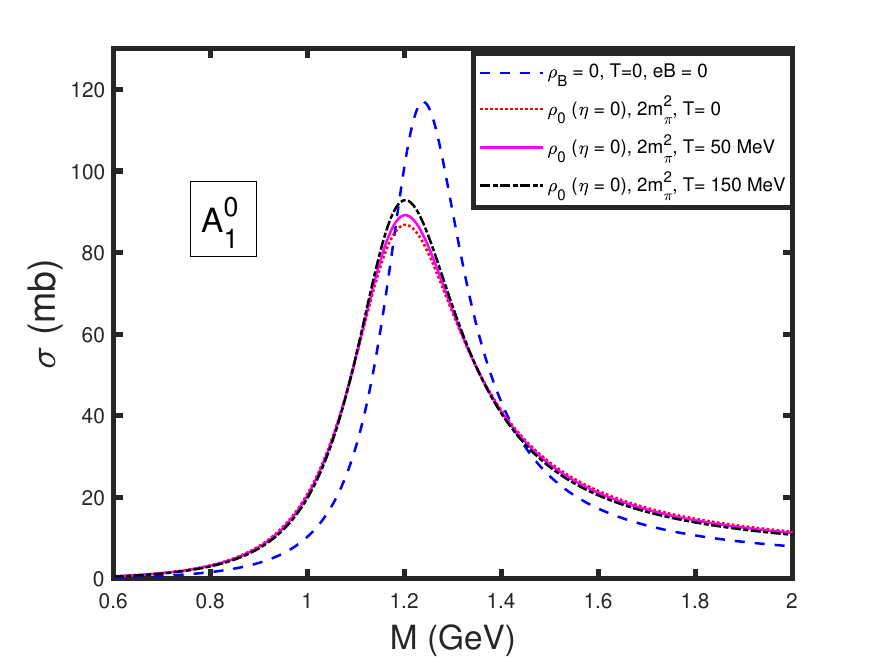} }}%
    \vspace{-0.4cm}
    \caption{\raggedright{ (a) Spectral function $A(M)$ (in Ge$V^{-1}$), (b) production cross-section $\sigma$ (in mb) of the neutral $A_1$ meson are plotted as functions of the invariant mass $M$ (in GeV) at $\rho_B=\rho_0$, in the symmetric nuclear matter ($\eta=0$), for $|eB|=2m_{\pi}^2$ at $T=0,\ 50$ and 150 MeV. The vacuum case with $\rho_B=0$, $T=0$ also $|eB|=0$ is shown for comparison as the dashed line. Effects of the magnetized Dirac sea with non-zero nucleonic AMMs are incorporated here.}}%
    \label{3}%
\end{figure}

In fig.\ref{2}, (a) the Breit-Wigner spectral functions $A(M)$ (in units of $GeV^{-1}$), and (b) the production cross-sections $\sigma$ (in units of $mb$) of $A_1^0$ are plotted as functions of the invariant mass $M$ (in units of GeV) in the symmetric nuclear matter at $\rho_0$ for $T=50$ MeV and various magnetic field strengths. The production cross-section of $A_1^0$ due to the scattering (and decay to) $\rho^{\pm}\pi^{\mp}$ channels are considered here. The hadronic decay widths of $A_1^0\rightarrow \rho^{\pm} \pi^{\pm}$ are computed using a phenomenological effective interaction vertex of $A_1\rho\pi$ as illustrated in Sec. \ref{sec4}. Effects of the magnetized Dirac sea along with the Landau energy levels contributions of protons and anomalous magnetic moments of the nucleons in hot and dense matter, are incorporated in the solutions of the scalar fields $\sigma$, $\zeta$, $\delta$ and $\chi$ within the framework of chiral effective model. The solutions are obtained at finite temperature (well below the critical temperature of chiral phase transition) and nuclear matter saturation density as well as zero baryon density in strong magnetic fields, which occurs at low density matter. As the $A_1$ mass changes very slightly with magnetic field as compared to the light vector meson $\rho$, there is almost no shift in the peak position of the spectral function at $\rho_0$ ($\eta=0$) with magnetic field. However, there is distinct difference in the peak position with change in baryon density $\rho_B$ from 0 to $\rho_0$. As mass decreases with density the peak shifted towards the low invariant mass region on the left in plot (a). In-medium decay widths increase with density, and also separately with magnetic field (for finite AMMs) at $\rho_0$. This is due to the fact that for non-zero anomalous magnetic moments of the nucleons, Dirac sea contribution in the hot magnetized matter leads to the effect of inverse magnetic catalysis at $\rho_0$ and $T=0,\ 50,\ 100, 150$ MeV in symmetric as well as asymmetric nuclear matter. As a consequence, the light quark condensates, the scalar gluon condensate and hence the masses decrease with magnetic field in the QCDSR approach. Therefore, from eq. (\ref{finaldecay}), in-medium partial decay widths of $A_1^0 \rightarrow \rho \pi$ increase with magnetic field. In our study the masses of the charged $\rho^{\pm}$ mesons are taken to be equal and the difference from the neutral partner $\rho^0$ is obtained from the scalar four quark condensate in the $c_3$ coefficient of the OPE side of QCDSR. In ref. \cite{k1}, the operators with non-zero spin dimension are also considered and due to the charge symmetry breaking effect, odd-dimensional operators in the OPE side lead to the non-degenerate mass of $K_1^+$ and $K_1^-$. The twist-2 operators are not considered here and the masses of the oppositely charged states are taken to be degenerate. An important thing to note here and in the subsequent plots is that the Dirac sea contribution in the magnetized matter is accounted for with non-zero anomalous magnetic moments of the nucleons. Fig.\ref{3} represents the plots of the (a) spectral functions and (b) productions cross-sections of $A_1^0$ at different temperatures $T=0,\ 50,\ 150$ MeV for a particular magnetic field of $2m_{\pi}^2$ in the symmetric nuclear matter at $\rho_0$. The vacuum case with $\rho_B=0$, $T=0$ and $|eB|=0$ is shown in the plots for comparison. It is observed that the production cross-section is increasing slightly with temperature as the decay width of $A_1\rightarrow \rho \pi$ process decreases and there is very small increment in the mass of the initial state particle of $A_1^0$ with $T$, at a fixed value of density $\rho_0$, and magnetic field strength. Although this amount of change in the spectral properties of $A_1$ meson is very less as compared to the change observed with respect to the variation in density and magnetic field including the Dirac sea effects. 
\begin{figure}%
    \centering
    \subfloat[\centering ]{{\includegraphics[width=8.4cm]{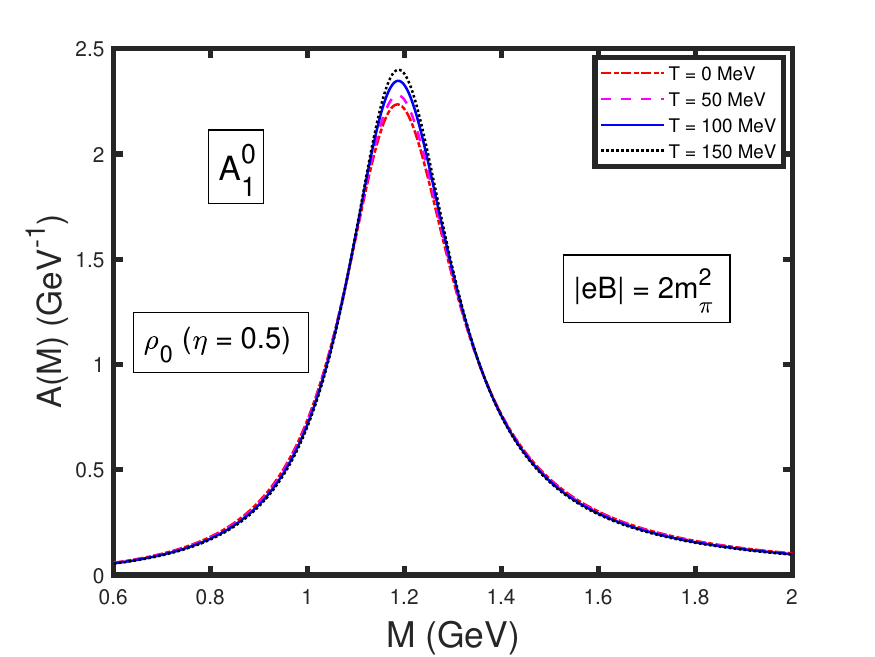} }}%
    \hspace{-0.8cm}
    \subfloat[\centering ]{{\includegraphics[width=8.4cm]{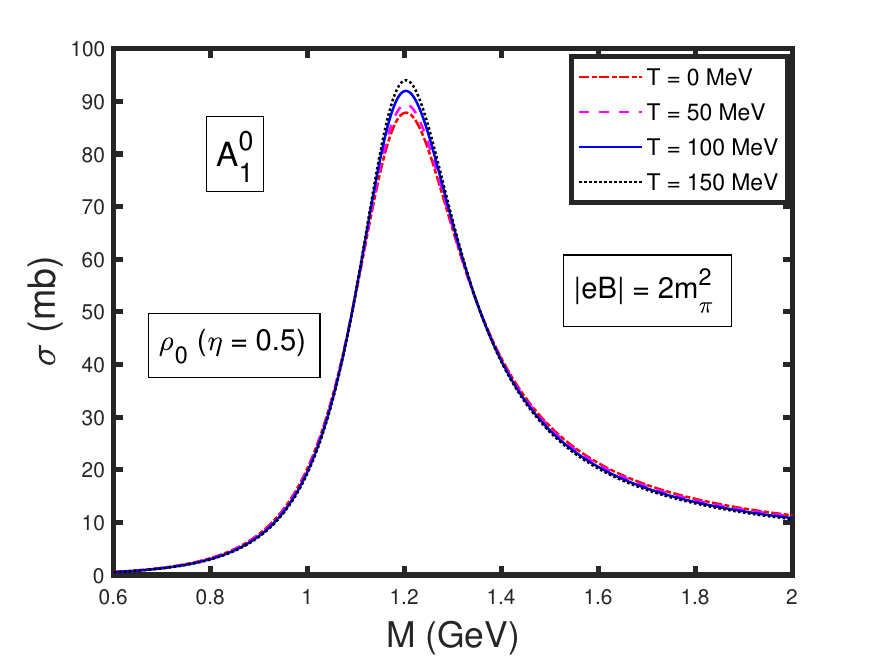} }}%
    \vspace{-0.4cm}
    \caption{\raggedright{ (a) Spectral function (b) production cross-section of the $A_1^0$ meson at $\rho_0$, in the asymmetric nuclear matter $\eta=0.5$, for $|eB|=2m_{\pi}^2$ at $T=0,\ 50,\ 100$ and 150 MeV. Effects of the magnetized Dirac sea is incorporated here for finite anomalous magnetic moments. }}%
    \label{4}%
\end{figure}
\begin{figure}%
    \centering
    \subfloat[\centering ]{{\includegraphics[width=8.4cm]{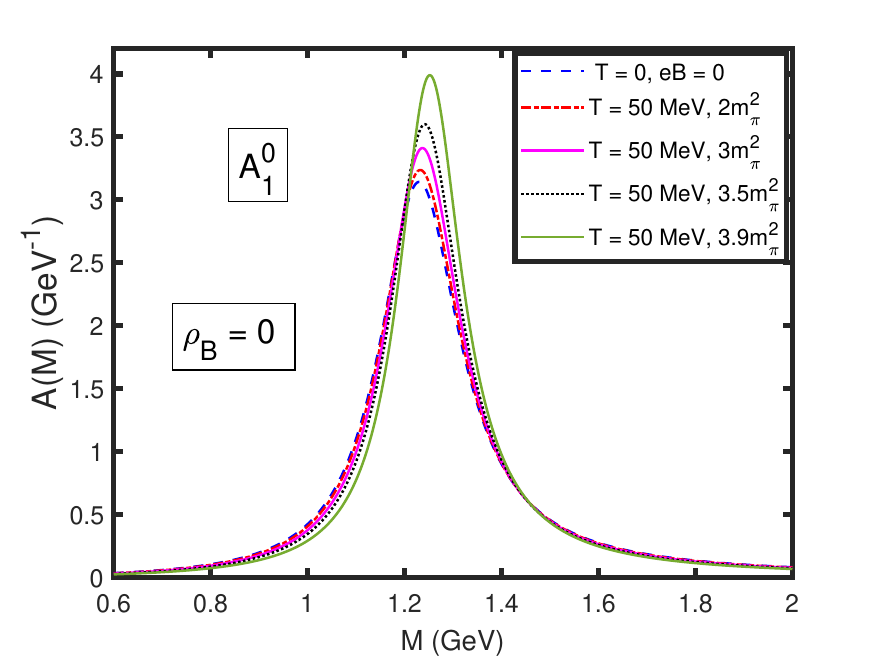} }}%
    \hspace{-0.8cm}
    \subfloat[\centering ]{{\includegraphics[width=8.4cm]{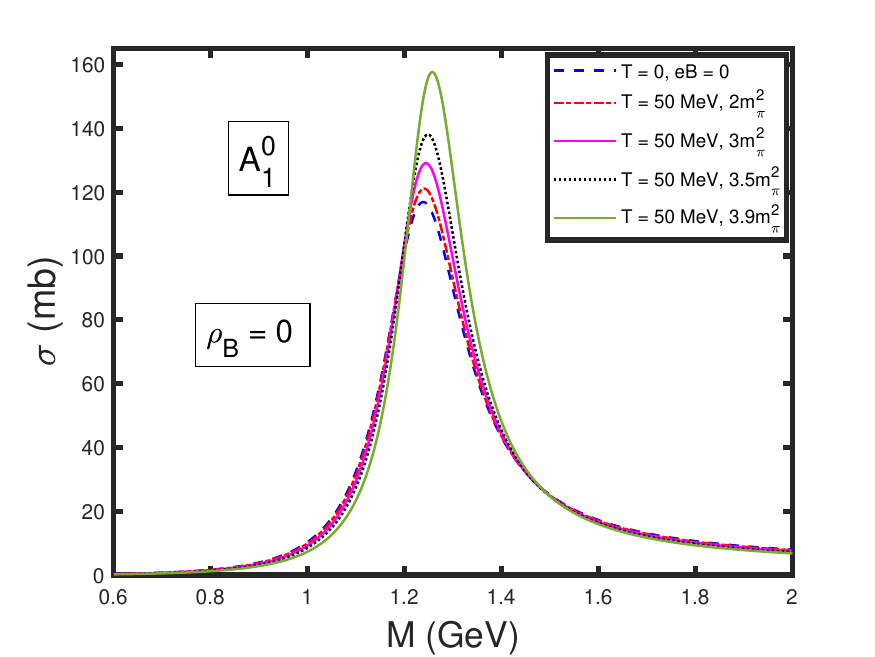} }}%
    \vspace{-0.4cm}
    \caption{\raggedright{ (a) Spectral function $A(M)$ (in Ge$V^{-1}$), (b) production cross-section $\sigma$ (in mb) of $A_1^0$ meson are plotted as functions of the invariant mass $M$ (in GeV) at $\rho_B=0$, $T= 50$ MeV for $|eB|/m_{\pi}^2 =2,\ 3,\ 3.5,\ 3.9$. The vacuum case with $\rho_B=0$, $T=0$ also $|eB|=0$ is shown for comparison as the dashed line. Effects of the magnetized Dirac sea with non-zero nucleonic AMMs are incorporated here.}}%
    \label{5}%
\end{figure}
\begin{figure}%
    \centering
    \subfloat[\centering ]{{\includegraphics[width=8.4cm]{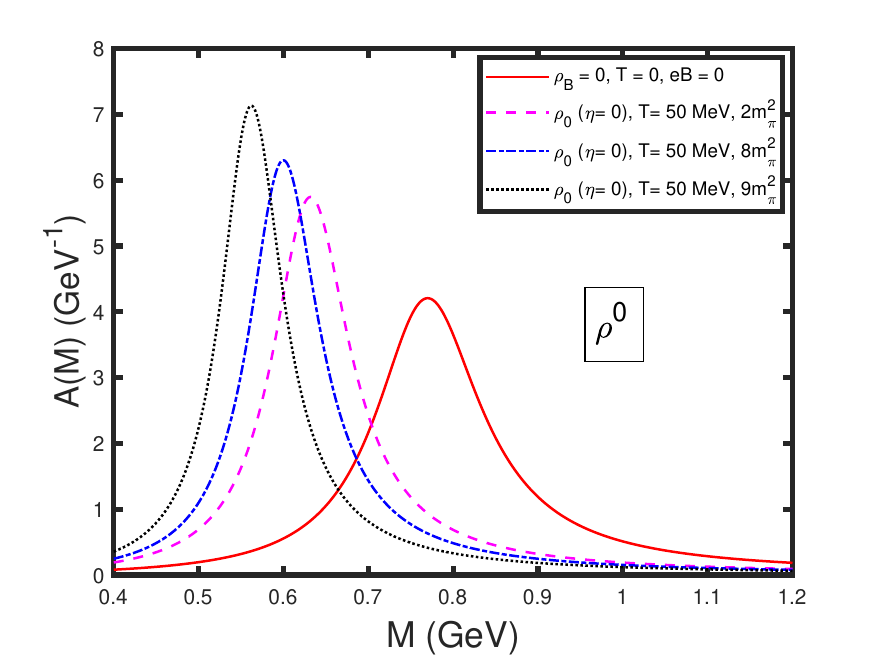} }}%
    \hspace{-0.8cm}
    \subfloat[\centering ]{{\includegraphics[width=8.4cm]{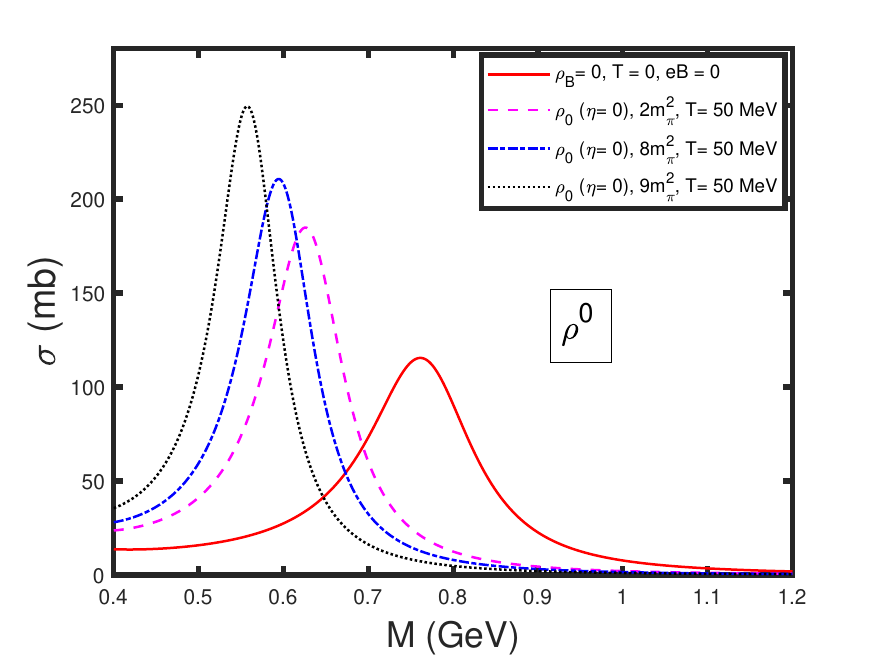} }}%
    \vspace{-0.4cm}
    \caption{\raggedright{ (a) Spectral function $A(M)$ (in Ge$V^{-1}$), (b) production cross-section $\sigma$ (in mb) of the neutral $\rho$ meson are plotted as functions of the invariant mass $M$ (in GeV) at $\rho_0$, $\eta=0$, and $T= 50$ MeV for $|eB|/m_{\pi}^2 =2,\ 8,\ 9$. The vacuum case with $\rho_B=0$, $T=0$ also $|eB|=0$ is shown for comparison as the solid line. Effects of the magnetized Dirac sea with non-zero nucleonic AMMs are incorporated here.}}%
    \label{6}%
\end{figure}

Fig.\ref{4} shows the similar plots as fig.\ref{3} for pure neutron rich matter with isospin asymmetry parameter $\eta=0.5$ at $\rho_0$. The production cross-section is observed to increase very slightly with temperature at fixed magnetic field strength of $2m_{\pi}^2$, accounting for the effects of the magnetized Dirac sea with finite AMMs. 

\begin{figure}%
    \centering
    \subfloat[\centering ]{{\includegraphics[width=8.4cm]{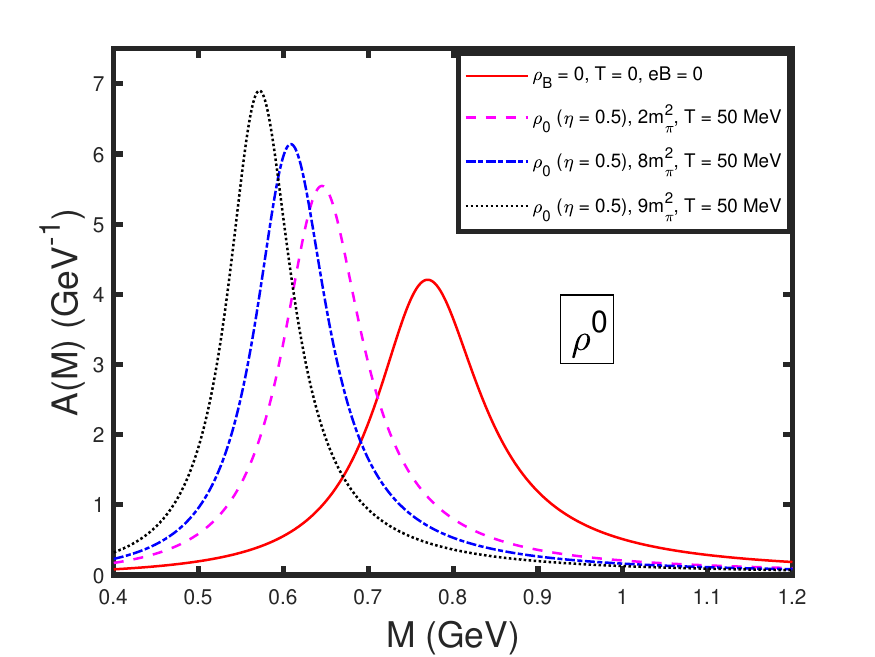} }}%
    \hspace{-0.8cm}
    \subfloat[\centering ]{{\includegraphics[width=8.4cm]{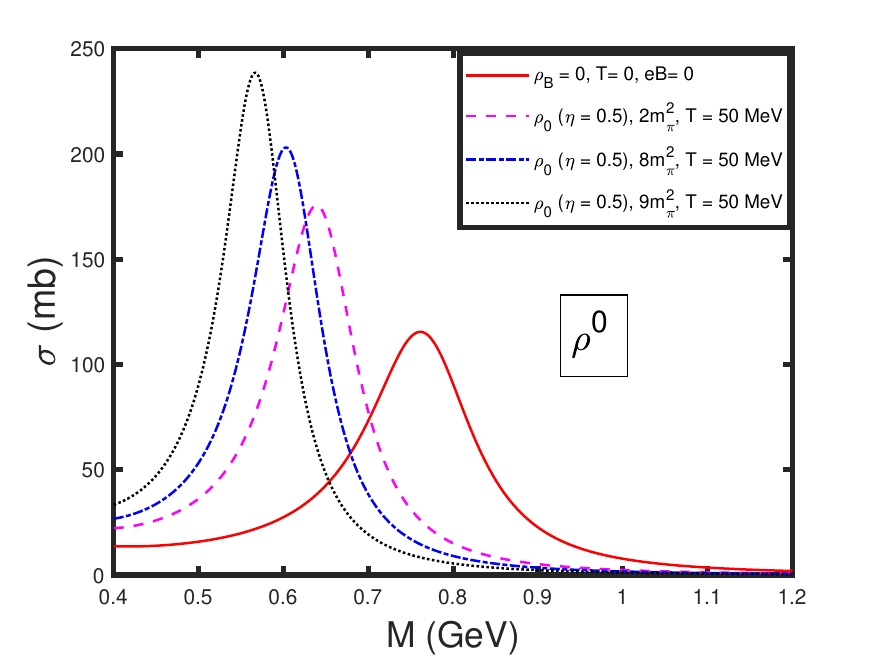} }}%
    \vspace{-0.4cm}
    \caption{\raggedright{ (a) Spectral function (b) production cross-section of $\rho^0$ meson are plotted as functions of the invariant mass $M$ (in GeV) at $\rho_0$, $\eta=0.5$, and $T= 50$ MeV for $|eB|/m_{\pi}^2 =2,\ 8,\ 9$. Rest of the description is similar as in fig.\ref{6}.}}%
    \label{7}%
\end{figure}

In fig.\ref{5}, similar plots of $A_1^0$ meson are shown at $\rho_B=0$, $T=50$ MeV, and for different values of magnetic field $|eB|/m_{\pi}^2=2,\ 3,\ 3.5$ and 3.9. The mass of $A_1^0$ meson rise with magnetic field at zero density and at any fixed temperature, due to the effects of magnetic catalysis for both non-zero and zero AMMs of the Dirac sea of nucleons. Hence the decay width in the $A_1\rightarrow \rho \pi$ channel decreases with $|eB|$ at a fixed temperature. This leads to an increment in $\sigma$ with magnetic field at $\rho_B=0$ and $T=50$ MeV. Thus, in the magnetized vacuum when there remains only the contribution of the Dirac sea, the production of $A_1^0$ meson is found to be increased due to the important phenomena of magnetic catalysis for both zero and non-zero nucleonic AMMs and at any temperature.   
In fig.\ref{6}, (a) the in-medium Breit Wigner spectral function $A(M)$, and (b) the production cross-section $\sigma$ are plotted as functions of the invariant mass $M$ for the neutral light vector meson $\rho$, at $\rho_B=\rho_0$, $\eta=0,$ and $T=50$ MeV for $|eB|/m_{\pi}^2=2,\ 8,\ 9$. The pure vacuum case in solid line is shown to emphasize on the in-medium spectral properties of $\rho^0$ meson due to the effects of inverse magnetic catalysis in hot and dense matter. The contributions of non-zero anomalous magnetic moments of the Dirac sea of nucleons are taken into account. Similar behavior are shown in fig.\ref{7} for the asymmetric magnetized nuclear matter at $\eta=0.5$. The production of $\rho^0$ mesons from the scattering of $\pi^+\pi^-$ mesons and also its decay to $\rho^0 \rightarrow\pi^+\pi^-$ channel is considered in the calculation of Breit Wigner spectral function from eq.(\ref{spec}) and cross-section from eq.(\ref{cross}). The mass of $\rho$ meson decreases with magnetic field at the nuclear matter saturation density, for both $\eta=0,\ 0.5$, and finite AMMs of the nucleons accounting for the effects of magnetized Dirac sea, as it is obvious from plot (b) in fig.(\ref{1}). The contributions of Fermi distribution functions in the number and scalar densities of nucleons incorporate the effects of temperature in the current study. The decreasing nature in mass with $|eB|$ for non-zero AMMs of the nucleons, is obtained for different values of temperatures $T=0,\ 50,\ 100$ and 150 MeV. From eq.(\ref{rhodecay}) the in-medium decay width of $\rho \rightarrow \pi^+\pi^-$ process is calculated using an effective Lagrangian of $\rho\pi$ interaction with the experimentally fitted coupling parameter $\tilde{g}=6.05$ \cite{klingl2}. The modified decay width is obtained from the in-medium mass of the parent particle $\rho$ meson with no medium changes in $\pi$ mesons masses in our present study, as stated before. The decay width is seen to decrease with magnetic field at the given condition from eq.(\ref{rhodecay}). Hence, the production cross-section of $\rho\rightarrow \pi^+\pi^-$ process increases with $|eB|$ due to the phenomena of inverse magnetic catalysis at $\rho_0$ and for finite AMMs of the nucleons. However, unlike the case of $A_1$ meson, the mass of $\rho$ changes considerably with magnetic field in this process and the shift in the peak position is significantly visible towards the low invariant mass region of $\rho$ meson. In ref.\cite{chernodub}, the minimal effective mass squared of the free charged, spin-0 pions $\pi^{\pm}$ received a positive contribution due to the lowest Landau energy level contribution in an external magnetic field background, which is also known as point-article correction. It has been used to study the onset of $\pi^{\pm}$ stability regions leading to the longer lifetime of $\rho$ mesons by closing the main strong decay channels of $\rho\rightarrow \pi^{+}\pi^-$. 

\begin{figure}%
    \centering
    \subfloat[\centering ]{{\includegraphics[width=8.4cm]{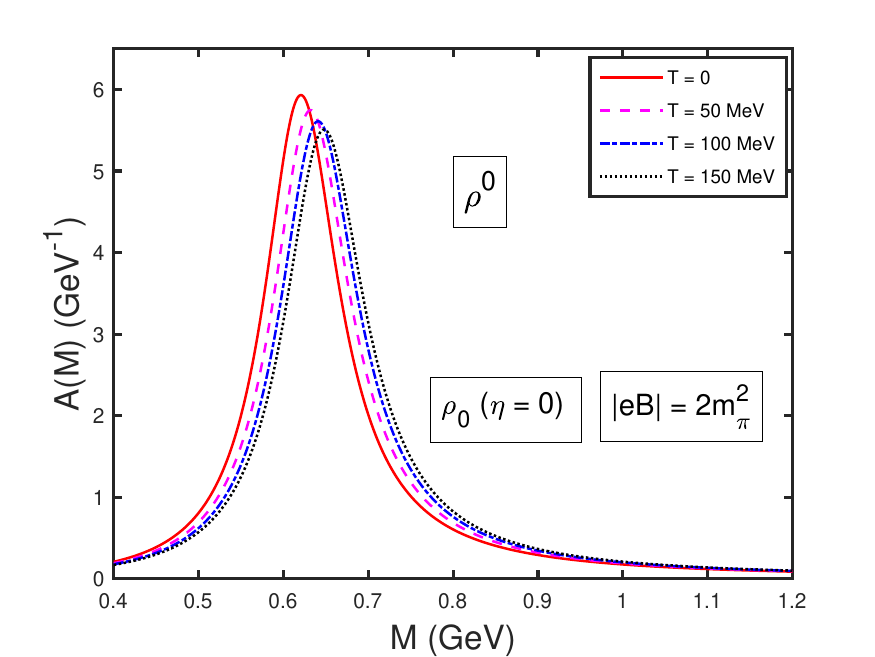} }}%
    \hspace{-0.8cm}
    \subfloat[\centering ]{{\includegraphics[width=8.4cm]{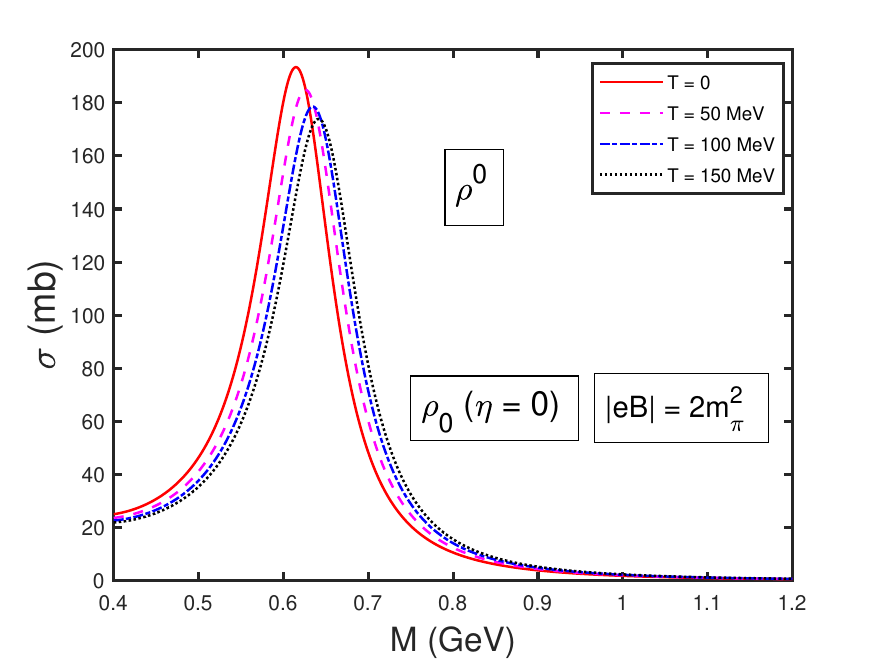} }}%
    \vspace{-0.4cm}
    \caption{\raggedright{ (a) Spectral function (b) production cross-section of $\rho^0$ meson are plotted as functions of the invariant mass $M$ (in GeV) at $\rho_0$, $\eta=0$, and $T= 0,\ 50,\ 100,\ 150$ MeV for $|eB| =2m_{\pi}^2$. Effect of magnetized Dirac sea is incorporated with finite nucleonic AMMs.}}%
    \label{8}%
\end{figure}
\begin{figure}%
    \centering
    \subfloat[\centering ]{{\includegraphics[width=8.4cm]{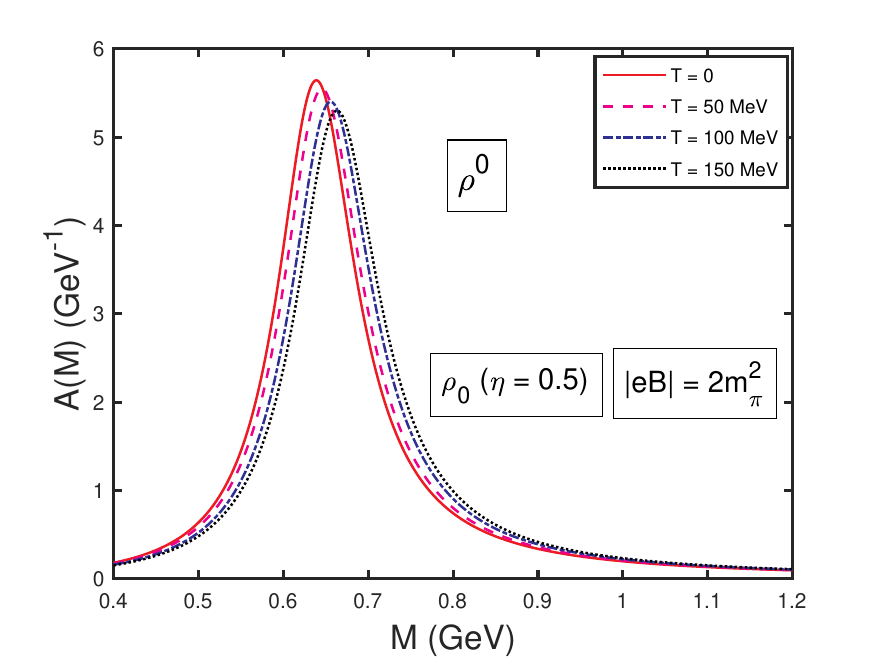} }}%
    \hspace{-0.8cm}
    \subfloat[\centering ]{{\includegraphics[width=8.4cm]{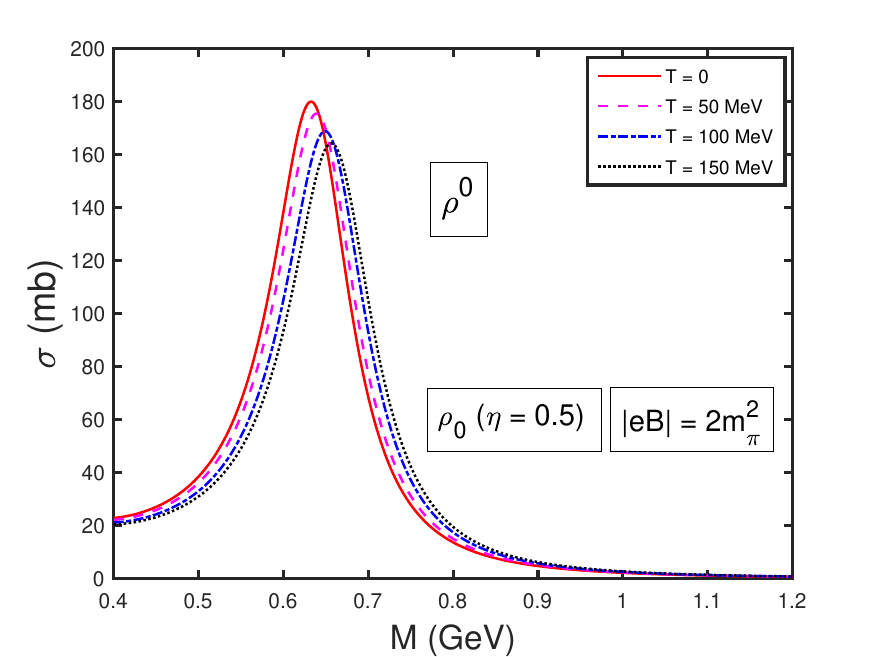} }}%
    \vspace{-0.4cm}
    \caption{Similar as in fig.\ref{8} for asymmetric nuclear matter at $\eta=0.5$.}%
    \label{9}%
\end{figure}

\begin{figure}%
    \centering
    \subfloat[\centering ]{{\includegraphics[width=8.4cm]{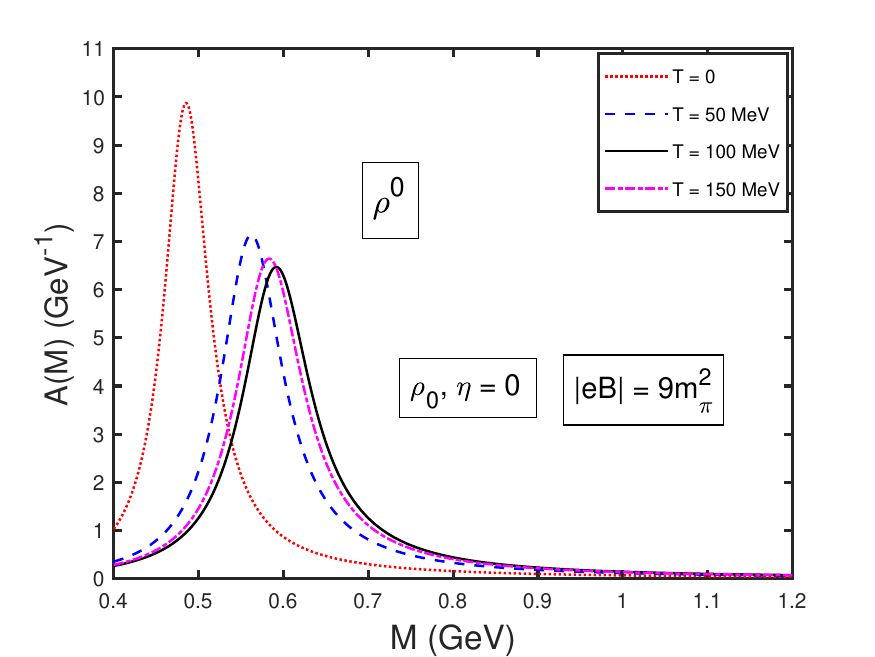} }}%
    \hspace{-0.8cm}
    \subfloat[\centering ]{{\includegraphics[width=8.4cm]{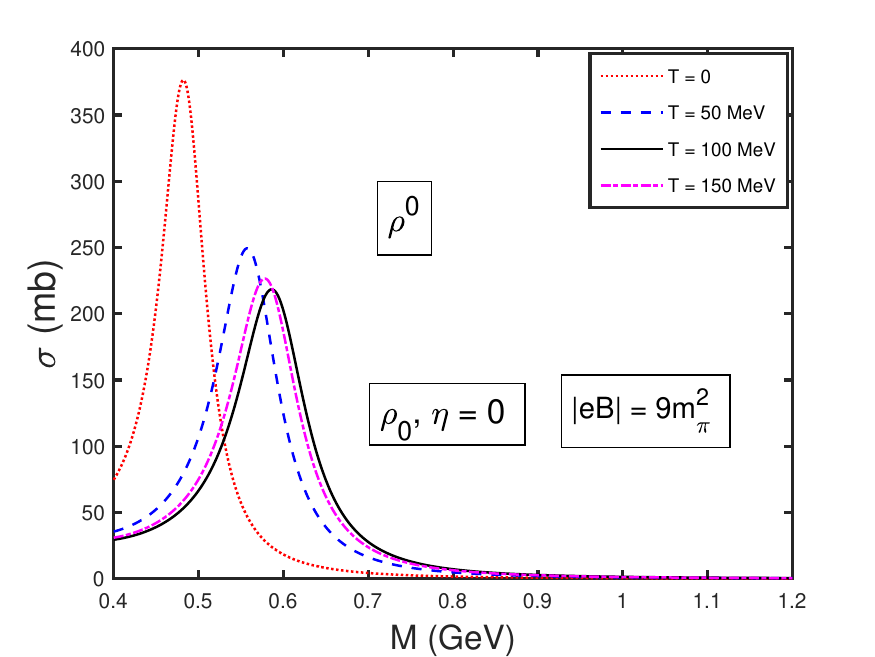} }}%
    \vspace{-0.4cm}
    \caption{\raggedright{ (a) Spectral function (b) production cross-section of $\rho^0$ meson are plotted as functions of the invariant mass $M$ (in GeV) at $\rho_0$, $\eta=0$, and $T= 0,\ 50,\ 100,\ 150$ MeV for $|eB| =9m_{\pi}^2$. Effects of inverse magnetic catalysis is incorporated. }}%
    \label{9.1}%
\end{figure}

\begin{figure}%
    \centering
    \subfloat[\centering ]{{\includegraphics[width=8.4cm]{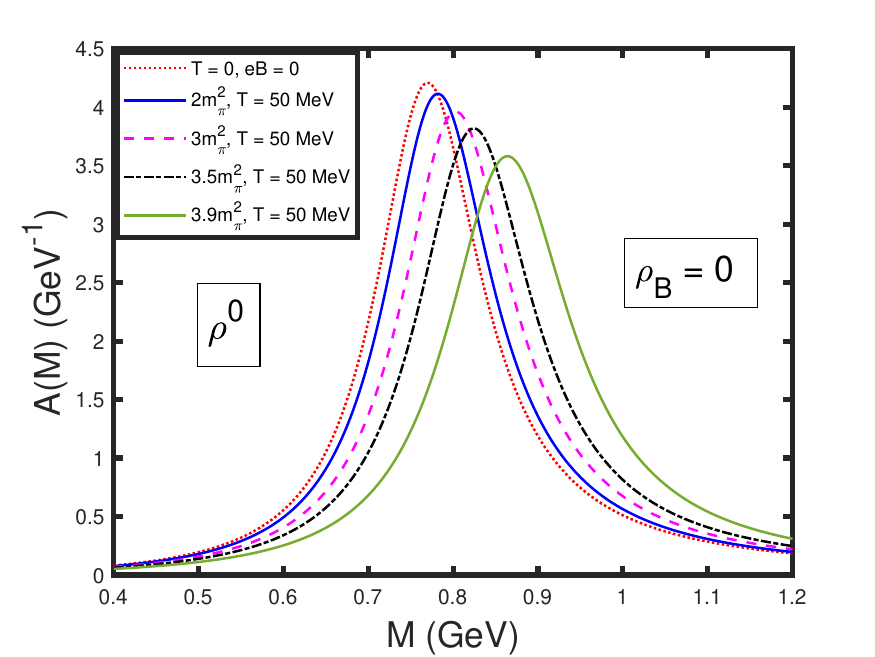} }}%
    \hspace{-0.8cm}
    \subfloat[\centering ]{{\includegraphics[width=8.4cm]{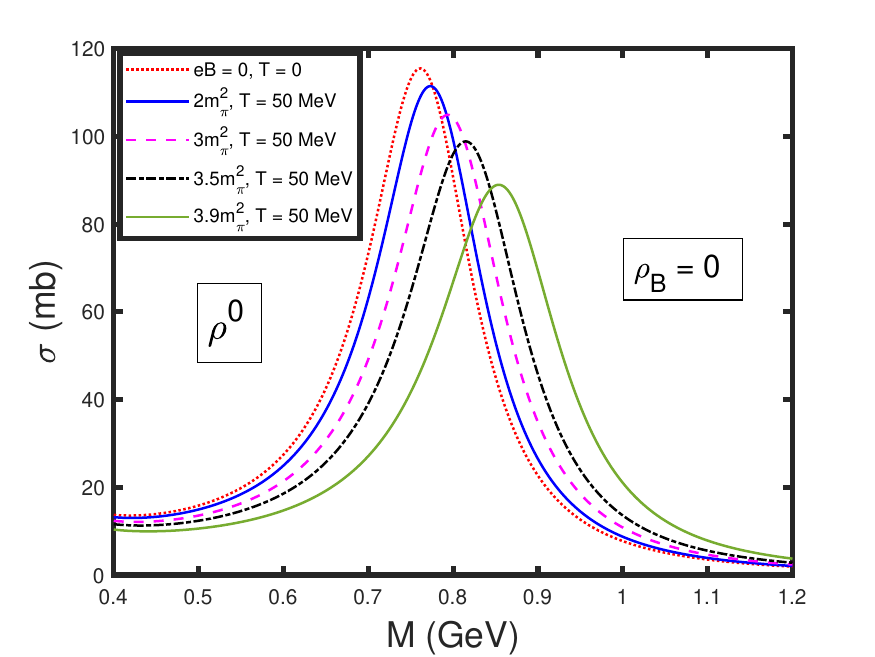} }}%
    \vspace{-0.4cm}
    \caption{\raggedright{(a) Spectral function (b) production cross-section of $\rho^0$ meson are plotted as functions of the invariant mass $M$ (in GeV) at $\rho_B=0$, and $T= 50$ MeV for $|eB|/m_{\pi}^2 =$ $2,\ 3,\ 3.5$, and 3.9. The pure vacuum case is shown for comparison as the dotted line. Effects of the magnetized Dirac sea are incorporated with finite nucleonic AMMs.}}%
    \label{10}%
\end{figure}
\begin{figure}%
    \centering
    \subfloat[\centering ]{{\includegraphics[width=8.4cm]{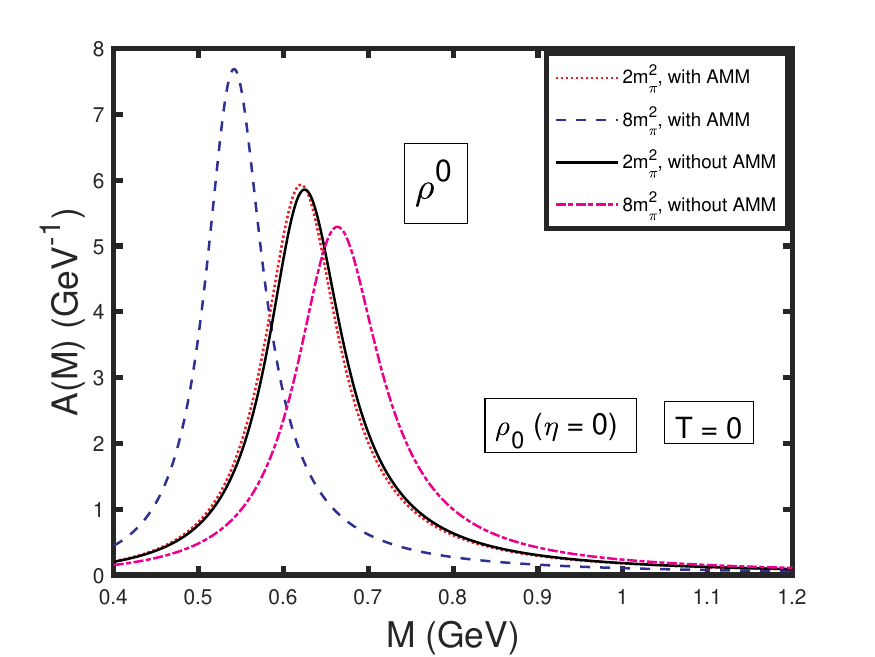} }}%
    \hspace{-0.8cm}
    \subfloat[\centering ]{{\includegraphics[width=8.4cm]{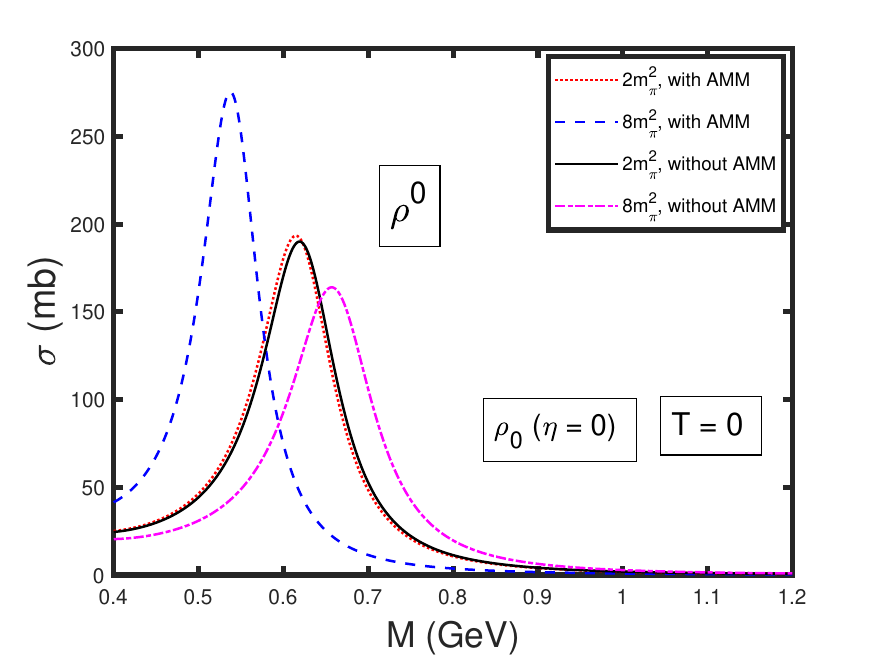} }}%
    \vspace{-0.4cm}
    \caption{\raggedright{(a) Spectral function (b) production cross-section of $\rho^0$ meson are plotted as functions of the invariant mass $M$ (in GeV) at $\rho_0$, $\eta=0$ and $T= 0$ MeV. The effects of the magnetized Dirac sea for the case of finite anomalous magnetic moments and zero AMM are shown for comparison for $|eB|=2m_{\pi}^2$ and $8m_{\pi}^2$. Effects of the magnetized Dirac sea are incorporated with finite nucleonic AMMs.}}%
    \label{11}%
\end{figure}

\begin{figure}%
    \centering
    \subfloat[\centering ]{{\includegraphics[width=8.4cm]{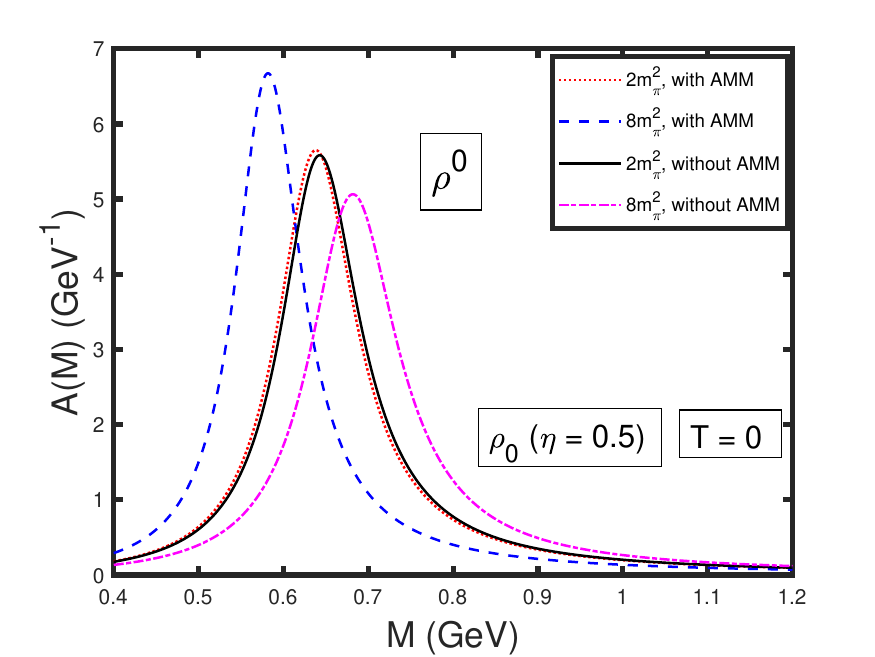} }}%
    \hspace{-0.8cm}
    \subfloat[\centering ]{{\includegraphics[width=8.4cm]{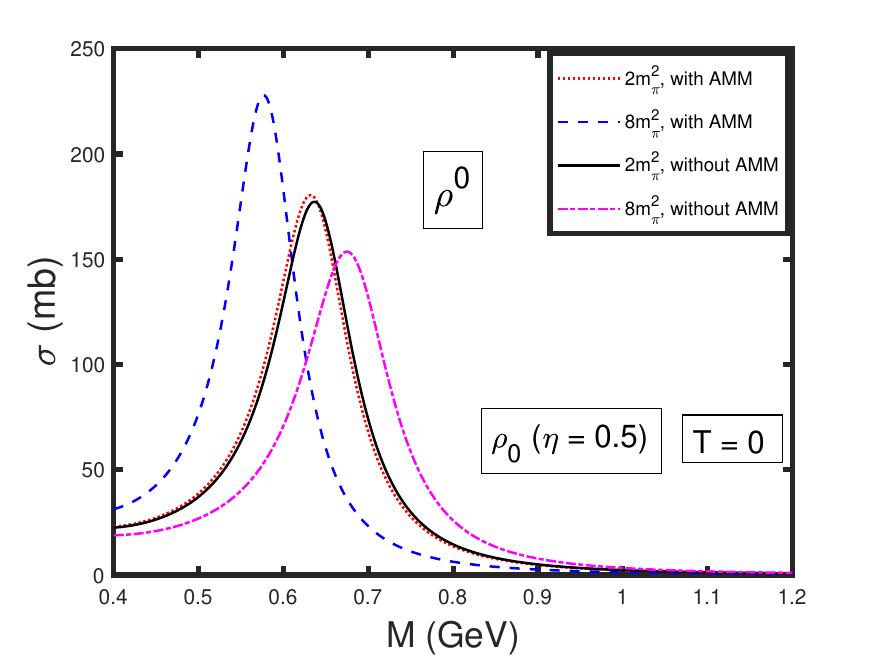} }}%
    \vspace{-0.4cm}
    \caption{{Same as in fig.\ref{11} for the asymmetric nuclear matter $\eta=0.5$.}}%
    \label{12}%
\end{figure}

Figs. \ref{8} and \ref{9} exhibit the plots of (a) $A(M)$ and (b) $\sigma$ of $\rho^0$ meson with variation in $M$ at the given values of density $\rho_0$, magnetic field $|eB|=2m_{\pi}^2$, for four temperatures $T=0,\ 50,\ 100,\ 150$ MeV in the symmetric ($\eta=0$) and asymmetric ($\eta=0.5$) nuclear matter, respectively. The effects of the magnetized Dirac sea with finite AMMs are incorporated in this behavior. At the given conditions, in-medium mass of $\rho$ meson increases very slightly with temperature at lower magnitude of magnetic fields e.g., at $2m_{\pi}^2$, leading to the increasing decay width of $\rho \rightarrow \pi^+\pi^-$ channel by eq.(\ref{rhodecay}). As a result, the production cross-section of the neutral $\rho$ meson decreases slowly with temperature in plot (b) at the nuclear matter saturation density. The peak position of the spectral function also shifts gradually towards the high invariant mass region with temperature in plot (a).

In fig.\ref{9.1}, similar plots as fig.\ref{8} are shown at the magnetic field strength of $9m_{\pi}^2$. Here the peak position of the spectral function in plot (a) shifts slightly towards left at high enough temperature of $T=100$ to 150 MeV, and the production cross-section also increases slightly due to the similar variation of temperature. This behavior is obtained because of the lowering of light quark condensates at high temperature and high magnetic field, accounting for the effects of inverse magnetic catalysis. 

Fig.\ref{10} contains the spectral behavior of $\rho^0$ meson similar to the previous plots at zero baryon density $\rho_B=0$, $T=0$, for $|eB|/m_{\pi}^2=2,\ 3,\ 3.5,\ 3.9$. The phenomena of magnetic catalysis at $\rho_B=0$, for both finite and vanishing AMMs of nucleons, mass of $\rho$ meson increases with magnetic field accompanied by increased decay widths of $\rho \rightarrow \pi^+ \pi^-$ process. Hence, the production cross section $\sigma$ decreases with magnetic field in plot (b), and the peak of the spectral function in (a) is shifted towards the high $M$ region on the right.  

In figs. \ref{11} and \ref{12}, the effect of the magnetized Dirac sea with finite nucleonic AMMs, is compared to the case when nucleons' AMMs are taken to be zero at the nuclear matter saturation density $\rho_0$, at zero temperature, for symmetric ($\eta=0$) and asymmetric ($\eta=0.5$) nuclear matter, respectively. For non-zero AMMs, the Dirac sea contribution in the magnetized nuclear matter leads to the phenomena of inverse magnetic catalysis and the opposite effect of magnetic catalysis is obtained when the AMM is taken to be zero. Therefore, the mass of $\rho$ meson is seen to be increasing (decreasing) with magnetic field due to the phenomena of magnetic (inverse) catalysis at $\rho_0$ for both the conditions of $\eta=0$ and 0.5. In plot (a) of figs. \ref{11}-\ref{12}, the peak position of the spectral function is therefore shifted towards the low $M$ region (on the left) for the variation of magnetic field strength from $|eB|=2$ to $8m_{\pi}^2$ for finite anomalous magnetic moments of the nucleons, which then shifts towards the high $M$ region (on the right) for the case of without AMMs of nucleons. The variation of $\rho$ meson mass in this way leads to the increasing (decreasing) decay width of $\rho^0\rightarrow \pi^+\pi^-$ process due to the effects of magnetic (inverse) catalysis at $\rho_0$ without (with) anomalous magnetic moments of the nucleons in the magnetized Dirac sea contribution. As a consequence, the production cross-section of $\rho$ meson decreases (increases) with magnetic field in plot (b), corresponding to the case of zero (non-zero) anomalous magnetic moments of nucleons.

The vacuum spectral functions of the vector and axial-vector meson states are related to some physical processes e.g., scattering. The vectorial spectral function can be obtained from
the $e^+e^-$ annihilation into an even number of pions. While the axial-vector one can be extracted from data on $\tau$ decay into a $\nu_{\tau}$ and an odd-number of pions. Their vacuum spectral functions are different which gives one of the experimental signatures that the chiral symmetry is spontaneously broken. The presence of pions in the thermal heat bath can lead to the mixing between the vacuum vector and axial-vector spectral functions via a model independent "mixing theorem". To the lowest order in temperature there is obtained no change in the spectral shape itself. Only a temperature dependent coupling between the vector and axial-vector correlators can be obtained. % The transition temperature as obtained from lattice QCD data is 155 MeV, which agrees well with that of the virial expansion of mixing ansatz satisfying the Weinberg sum rules to be 160 MeV. Such low temperature expansion can not be expanded easily to high temperature region.
%In the context of chiral SU(3) model, the chiral transition temperature is around 150 MeV. 
However, as an experimental signal it is instructive to observe the consequences on the dilepton spectra or, the $l^+l^-$ invariant mass spectra. Dileptons or, $l^+l^-;\ l=e, \mu$ pairs are provided to be an important tool to measure the medium modifications of the vector mesons e.g., $\rho,\omega, \phi$ in nuclear media, as their final-state interactions are very negligible. Both the mixed vector and axial-vector correlator are indistinguishable down to 1 GeV, which is lower than its vacuum position at around 1.5 GeV. Hence the effects of finite temperature lead to the lowering of the duality threshold from its 
vacuum value of 1.5 GeV down to 1 GeV in medium. Therefore, the direct signature of chiral restoration is the degeneracy between vector and axial-vector spectral functions. As the vector current directly couples to the virtual photons which further decay into $l^+l^-$, the vector spectral function is directly related to the physically measurable observable. The broadening or melting of the resonances in the medium that flattens the $\rho$ meson resonance structure and generates low-mass dilepton enhancement below rho mass. 
The spontaneous breaking of chiral symmetry (SBCS) is inherently a soft phenomena occurring at scales of less than or around 1 GeV, for this the low mass dilepton spectra is considered to be one of the most promising observables to detect changes in the chiral properties. E.g., change in the condensates at finite temperature and/or density, also with magnetic field in the context of non-central heavy ion collisions.
%The low mass dilepton emissivity, governing the thermal radiation from a hot fireball, is directly proportional to the hadronic vector spectral function for e.g., of the $\rho$ meson.
At the hadronic level, SBCS is related to the splitting of opposite parity states in chiral multiples e.g., $\pi-\sigma$, $\rho-A_1$ etc. Thus, to extract the signatures of (or, the approach to) chiral symmetry restoration from dilepton spectra dominated by the $\rho$ channel, one is lead to the theoretical study of the in-medium spectral properties of $A_1$ meson as well. Phenomenologically it is also rather well established that the low mass dilepton enhancement is a common feature observed in every experiments, from the low energy SIS experiments with center-of-mass (square-root) $\sqrt{s}=2.25$ GeV \cite{sis} via SPS ($\sqrt{s}=8.8, 17.3$ GeV) \cite{sps}, which is again obtained via the RHIC beam energy scan ranging up to its maximum energy ($\sqrt{s}= 19.6, 27,39,62,200$ GeV) \cite{rhic}.
It shows a broad $\rho$ meson spectral function due to the thermal radiation mostly of hadronic origin.
Thus, utilizing the $\rho$ meson spectral function which has direct connection to the observed dilepton spectra, the spectral function of its chiral partner $A_1$ can be evaluated using the in-medium values of the condensates from a chiral effective model and using the constraints of QCD Sum rules. It is almost impossible to measure $A_1$ spectral function via dilepton emissions, as the neutral $A_1$ current can only couple to dileptons via weak interaction through $Z^0$ boson \cite{hayano}.  
However, one point to be noted here is that the chirally symmetric condensates do not need to be vanished at the critical point of the chiral phase transition. Hence, it is not straightforward to relate the spectral modifications of vector mesons to the chiral symmetry restoration. 

The amount of shift in the spectral peak position of $\rho^0$ meson, which has the direct connection to the dilepton pairs are determined in the present work incorporating the effects of the magnetized Dirac sea at finite temperature and density matter in a background magnetic field. The amount of shift in the peak position of the invariant mass spectra of $\rho^0$ meson spectral function $A(M)$ (in Ge$V^{-1}$) from its vacuum position of $M=0.77$ GeV is given in the symmetric nuclear matter $\eta=0$, at $\rho_0$ and $T=50$ MeV, for $|eB|=2,8,9m_{\pi}^2$ respectively as (in GeV) -0.138, -0.17, -0.208, respectively, which are towards the low-invariant mass region on the left side as shown in fig.\ref{6} (a). Now the shifts with varying temperature at a fixed magnetic field strength of $2m_{\pi}^2$ and $T=50,100,150$ MeV are (in GeV) 0.138, 0.129, 0.122, respectively, as shown in fig.\ref{8} (a) (towards higher $M$ region from its vacuum position). The similar shifts obtained for $9m_{\pi}^2$ as 0.208, 0.178, 0.187 GeV at $T=50,100,150$ MeV, respectively, as in fig.\ref{9.1} (a). The spectral changes in the vacuum are obtained by incorporating the effects of the magnetized Dirac sea as an external effect. At $\rho_B=0$, $T=50$ MeV, the peak position is shifted towards right from its vacuum position by an amount of 0.012, 0.033, and 0.094 GeV for $|eB|=2,3,3.9m_{\pi}^2$, respectively, in fig.\ref{10} (a), as a consequence of the phenomena of magnetic catalysis originated due to the magnetized Dirac sea of nucleons. The data of spectral peak shifts given above are obtained while taking into account the finite anomalous magnetic moments of the nucleons at $\rho_B=0,\rho_0$, for various temperatures and magnetic field strengths. The lowering of the peak position with increasing magnetic field at $\rho_0$ ($\eta=0$) is due to the effect of inverse magnetic catalysis, while that at vacuum due to catalysis, in both cases AMMs are non-zero. The effects of temperature is somewhat noticeable at higher magnetic field strength e.g., at $9m_{\pi}^2$ for finite nucleonic AMMs. The shifting happen in the opposite direction as one goes from $T=100$ to 150 MeV with $|eB|$ changing from 2 to $9m_{\pi}^2$. At higher $eB$, the effects of inverse magnetic catalysis lead the peak position to shift towards the low-mass region with increasing temperature, i.e., an apparent tendency towards the chiral symmetry restoration. 

In the peripheral ultra-relativistic heavy ion collision experiments e.g., at RHIC with $\sqrt{s}=200$ GeV in Au+Au collisions and at LHC with $\sqrt{s}=2.76$ TeV in Pb+Pb collisions,  the early stage time evolution of the produced electromagnetic fields (at the origin) have been studied for the central as well as non-central collisions \cite{deng}. The contributions of the electrically charged participants, spectators, and remnant particles are taken into account. At a later stage of the collisions, e.g., at around $1 $ fm/c, the contributions are expected to come predominantly from the remnants as they move at a slower rate than the spectators from the collision zone. Studies have been discussed that the remnants may lead to the slowing down of the decay process of transverse electromagnetic (EM) fields. Although there might have been enhanced longitudinal components of EM fields due to the fluctuations of the remnants from the equilibrium position. For central collisions, any non-vanishing field is generated by the position fluctuations of the charged particles, which cause a sizable amount of fields in the transverse directions at the initial time which decay rapidly (to a large extent for higher center-of-mass energies). On the contrary, the off-central collisions lead to a dominant field component perpendicular to the reaction plane coming due to the motion of the spectators. However, the dominance of remnants over the spectators at a later time gives rise to the equal contribution to the magnitude of the transverse components of the magnetic field. From an estimate of the time-evolution of EM fields before and after collisions at the center-of-mass energy of RHIC, there can be obtained very small but a sizable component of the transverse fields at or around 1 fm/c \cite{deng}. Therefore, as studied in our present work the effects of the magnetic field on the mesons in a magnetized matter are coming out to be significant when incorporating the effects of the magnetized Dirac sea into the computation of mesons spectral properties. For the $A_1$ meson there is obtained a noticeable amount of modifications in the production cross sections, whereas there is obtained distinct peak shifts of $\rho$ meson spectral function with changing magnetic field strength at various other medium conditions as described in this section.

\section{Summary}
\label{sec7}
In the summary, we have investigated the in-medium spectral properties of the light vector $\rho$, $\omega$ and the light axial-vector $A_1$ mesons in the isospin asymmetric magnetized nuclear matter, accounting for the effects of (inverse) magnetic catalysis at zero and finite temperatures. The in-medium masses of $\rho$, $\omega$ and $A_1$ mesons are calculated within the QCD sum rule framework, using the Borel transform followed by the finite energy sum rules. The values of the other spectral parameters $F,\ s_0$, are modified by the medium effects. Masses are obtained in terms of the in-medium values of the light quark (up to the scalar four-quark condensate) and the scalar gluon condensates. The condensates are further obtained within the chiral effective model in terms of the scalar isoscalar fields non-strange $\sigma$, strange $\zeta$, scalar isovector $\delta$ and the scalar dilaton field $\chi$, at the given values of density, isospin asymmetry, magnetic field and temperature. The effects of the magnetized Dirac sea is incorporated through one-loop self energy functions of the nucleons due to their interactions with the scalar mesons $\sigma$, $\zeta$, and $\delta$ within the chiral $SU(3)$ model framework. However, the effects of finite anomalous magnetic moments of the Dirac sea of nucleons are incorporated to the self energy part by computing the weak field expansion (up to second order in $|\textbf{eB}|$) of the fermionic propagators. Effects of Landau energy levels of protons and anomalous magnetic moments of the Fermi sea of nucleons are also incorporated in the study of magnetized nuclear matter at finite temperature. Temperature effects are taken into account by introducing the Fermi distribution functions to the number ($\rho_{p,n}$) and scalar ($\rho^s_{p,n}$) densities of nucleons in solving the coupled equations of motion of the four scalar fields $\sigma$, $\zeta$, $\delta$ and $\chi$. The hadronic decay widths of $A_1\rightarrow \rho\pi$ channel is determined using an effective interaction Lagrangian of $A_1\rho\pi$ vertex from the in-medium masses of the $A_1^0$ and $\rho^{\pm}$ states by keeping the pion $\pi^{\pm}$ mass fixed. The in-medium decay widths in the main decay channel of $\rho^0\rightarrow \pi^+\pi^-$ is also determined using an effective Lagrangian of $\rho\pi$ interaction. In both cases the coupling parameter is fitted from the experimentally observed decay widths (or, decay branching ratio of several modes) to calculate the medium modified strong decay widths of neutral $\rho$ and $A_1$ mesons. From the in-medium masses and decay widths thus obtained, the spectral functions and production cross-sections of the neutral $A_1$ and $\rho$ mesons are investigated at different values of density, isospin asymmetry, temperature and magnetic field in addition to the important effects of (inverse) magnetic catalysis. The (decrement) increment in the values of the light quark condensates with magnetic field give rise to the phenomenon of (inverse) magnetic catalysis for (non-zero) zero anomalous magnetic moments of the nucleons at the nuclear matter saturation density $\rho_0$, for both $\eta=0$ and 0.5, and at $T=0,\ 50,\ 100,$ and 150 MeV of temperatures. The contribution of magnetic field at finite density matter is significant through the magnetized Dirac sea along with the additional important effects due to the non-zero anomalous magnetic moments of the Dirac sea of nucleons. There are observed to be non-trivial effects of temperature on the in-medium mass, decay widths, hence on the spectral function and production cross-sections of the light vector and axial-vector mesons. The masses of $\rho$ and $A_1$ mesons increase slightly with temperature at $\rho_0$ for finite nucleonic AMMs and magnetic field strength of $2m_{\pi}^2$, which show opposite behavior at high temperature region for $9m_{\pi}^2$, due to the effects of inverse magnetic catalysis. The productions of the light vector and axial-vector mesons are thus modified considerably due to the effects of (inverse) magnetic catalysis in a magnetized nuclear matter considered here at finite temperature case. These results may affect the experimental observables in the peripheral ultra relativistic heavy-ion collision experiments, where huge magnetic fields are expected to be produced at the initial stage of collisions with very high temperature.
 
\acknowledgements
Amruta Mishra acknowledges financial support from Department of Science and Technology (DST), Government of India (project no. CRG/2018/002226). Pallabi Parui wants to acknowledge Sourodeep De for many useful suggestions and a thoughtful discussion.

\end{document}